\documentclass[english,authoryear]{elsarticle}
\usepackage[T1]{fontenc}
\usepackage[latin9]{inputenc}
\usepackage{color}
\usepackage{bm}
\usepackage{amsmath}
\usepackage{amssymb}
\usepackage{graphicx}
\usepackage{setspace}
\usepackage{esint}
\makeatletter

\newcommand*\LyXZeroWidthSpace{\hspace{0pt}}

\journal{International Journal of Psychophysiology}

\usepackage{babel}

\makeatother

\usepackage{babel}
\usepackage{listings}
\begin{document}

\begin{frontmatter}{}

\title{The Surface Laplacian Technique in EEG: Theory and Methods}

\author[claudio]{Claudio Carvalhaes\corref{cor1}}

\ead{claudioc@stanford.edu}

\author[acacio]{J. Acacio de Barros}

\cortext[cor1]{Corresponding author}

\address[claudio]{Center for the Study of Language and Information, 220 Panama St,
Stanford, CA 94305, USA}

\address[acacio]{Liberal Studies, San Francisco State University, 1600 Holloway Ave,
San Francisco, CA 94312}
\begin{abstract}
This paper reviews the method of surface Laplacian differentiation
to study EEG. We focus on topics that are helpful for a clear understanding
of the underlying concepts and its efficient implementation, which
is especially important for EEG researchers unfamiliar with the technique.
The popular methods of finite difference and splines are reviewed
in detail. The former has the advantage of simplicity and low computational
cost, but its estimates are prone to a variety of errors due to discretization.
The latter eliminates all issues related to discretization and incorporates
a regularization mechanism to reduce spatial noise, but at the cost
of increasing mathematical and computational complexity. These and
several others issues deserving further development are highlighted,
some of which we address to the extent possible. Here we develop a
set of discrete approximations for Laplacian estimates at peripheral
electrodes and a possible solution to the problem of multiple-frame
regularization. We also provide the mathematical details of finite
difference approximations that are missing in the literature, and
discuss the problem of computational performance, which is particularly
important in the context of EEG splines where data sets can be very
large. Along this line, the matrix representation of the surface Laplacian
operator is carefully discussed and some figures are given illustrating
the advantages of this approach. In the final remarks, we briefly
sketch a possible way to incorporate finite-size electrodes into Laplacian
estimates that could guide further developments. \end{abstract}

\end{frontmatter}{}

\section{Introduction}

The surface Laplacian technique is a powerful method to study EEG.
It emerged back in the 1970's, from the seminal works of \citet{nicholson_theoretical_1973},
\citet{Freeman1975}, \citet{nicholson_theory_1975}, and \citet{hjorth_-line_1975}.
These works were followed by efforts to develop better computational
methods \citep[see][]{Gevins1988,gevins_dynamic_1989,Gevins1990,perrin_spherical_1989,law_high-resolution_1993,yao_high-resolution_2002,Carvalhaes2011}
as well as attempts to combine the surface Laplacian with other methods
\citep{Kayser2006a,Kayser2006,Carvalhaes2009}, making the technique
increasingly popular among EEG researchers. For instance, modern applications
include studies on generators of event-related potentials \citep{Kayser2006a,Kayser2006},
quantitative EEG \citep{Tenke2011}, spectral coherence \citep{Srinivasan2007,Winter2007},
event-related synchronization/desynchronization \citep{DelPercio2007},
phase-lock synchronization \citep{Doesburg2008}, estimation of cortical
connectivity \citep{Astolfi2007}, high-frequency EEG \citet{Fitzgibbon2013},
and brain-computer interface \citep{Lu2013}, just to mention a few. 

The motivation for the use of the surface Laplacian technique is grounded
on Ohm's law. This law establishes a local relationship between the
surface Laplacian of scalp potentials and the underlying flow of electric
current caused by brain activity \citep[see Appendix in][]{Carvalhaes2014}.
Presumably, this local relation should improve spatial resolution,
reflecting electrical activity from a more restricted cortical area
than what is observed in conventional topography. 

In contrast to other high-resolution techniques, such as cortical
surface imaging \citep{nunez_theoretical_1994,Yao1996,Yao2001}, the
surface Laplacian has the advantage of not requiring a volume conductor
model of the head or a detailed specification of neural sources, but
objections to it may arise due to a combination of factors. First,
there is a noticeable difficulty in understanding the technique in
depth \citep{nunez1994}. While EEG potentials can be physically understood
in terms of current flow or in analogy with the gravitational potential,
interpreting the surface Laplacian operation seems much less intuitive,
as it involves the application of a second-order partial differential
operator to a scalar potential distribution. Second, it appears surprising
to many that one can actually obtain a reference-independent quantity
from a signal which is seen to have been contaminated by a reference
in its very origin \citep[ch 7]{nunez_electric_2006}. Third, it is
not possible to guarantee that the theoretical advantages associated
to the use of the Laplacian differentiation will be preserved in the
practical world of imperfect measurements and finite samples. In fact,
reliable estimates of the Laplacian derivation are technically challenging,
and computational methods to perform this task are still a subject
of research.

The literature dedicated to provide theoretical explanations about
the surface Laplacian technique is scarce, but it contains substantial
and valuable information \citep[see][]{nunez_electric_2006,Tenke2012}.
Nevertheless, there are still some gaps that, if filled, can help
comprehension and permit a more intuitive view of the technique. It
is the primary intention of this paper to contribute to this matter.
For this purpose, Section\,\ref{sec:Theory} provides physical insights
on the surface Laplacian technique that are often not discussed in
the literature, but that are at the heart of some of the issues of
interest to the EEG researcher. Section\,\ref{sec:Computational-Aspects/Methods}
focuses on the computational aspects of this technique, providing
a comprehensive review of selected computational methods and their
advantages and limitations. Two sections were devoted to topics that
in our view deserve a more detailed discussion. In Section \ref{sec:The-L-Matrix}
we present the method of estimating the surface Laplacian derivation
by means of a matrix transformation, and Section \ref{sec:Data-Smoothing}
focuses on the regularization problem of smoothing splines that significantly
affects surface Laplacian estimates.

\section{An Overview of the Physics of EEG\label{sec:Theory}}

In this Section we discuss the main theoretical ideas behind the use
of the surface Laplacian. Our goal here is to introduce some concepts
from the physics of EEG that are often unfamiliar to the EEG researcher,
and that have direct relevance to the surface Laplacian technique.
We do not attempt to reproduce the main arguments about the relationship
between the surface Laplacian and the dura-surface potential, nor
with the Current Source Density (CSD), which are found, for example,
in \citet{nunez_electric_2006} and \citet{Tenke2012}. Instead we
focus on the more fundamental aspects that are often overlooked.

\subsection{Physical interpretation of the surface Laplacian\label{sub:Physical-Interpretation}}

To better understand the physical meaning of the surface Laplacian,
it is instructive to start with the volume Laplacian, or simply Laplacian,
and its relationship to the quasi-static electric field and potential.
Here our treatment is very standard, and can be found in more details
in any textbook on electricity and magnetism, such as \citet{jackson_classical_1999}
or \citet{schey_div_2004}, and the reader interested in more detail
is referred to them. Our discussion will focus only on aspects that
are directly relevant to EEG.

When we talk about scalp EEG, we are always referring to the potential
with respect to some reference electrode. But, from a physics point
of view, the \emph{electric field} is more fundamental than the potential.
To see this, let us recall that the field has a measurable effect:
for a particle of charge $q$, the field $\mathbf{E}(\mathbf{r})$
at the position $\mathbf{r}$ is operationally \emph{defined} by the
ratio between the electric force $\mathbf{F}_{E}$ acting on the charge
at $\mathbf{r}$ and the value of $q$. Here we use the standard notation
of representing scalar quantities in italics and vectors in boldface,
e.g. $q$ for the electric charge and $\mathbf{E}$ for the electric
field, and we also use the symbol ``$\equiv$'' for an equality
that arises from a definition. From this definition, the field is
a quantity that measures the force per unit of charge, i.e., $\mathbf{E}$
(as a function of the position $\mathbf{r}$) is given by 
\begin{equation}
\mathbf{E}(\mathbf{r})\equiv\frac{1}{q}\mathbf{F}_{E}(\mathbf{r}).
\end{equation}
For simplicity, we are ignoring any temporal dependence of $\mathbf{E}$,
but the results remains the same, for all practical purposes, in the
typical range of frequencies involved in the brain electrical activity
\citep{jackson_classical_1999,nunez_electric_2006}.

Electric fields generated by point charges are described by Coulomb's
Law. A consequence of this law is that $\mathbf{E}$ is a conservative
field, i.e., the line integral 
\begin{equation}
V_{AB}=-\int_{\mathbf{r}_{A}}^{\mathbf{r}_{B}}\mathbf{E}(\mathbf{r})\cdot d\mathbf{r}\label{eq:line-integral-field}
\end{equation}
between two points $\mathbf{r}_{A}$ and $\mathbf{r}_{B}$ is independent
of the path along which it is computed. The value $V_{AB}$ has an
important physical meaning: it corresponds to the work per unit of
charge necessary to be done on a charged particle when it goes through
the field \textbf{$\mathbf{E}$ }from position $\mathbf{r}_{A}$ to
$\mathbf{r}_{B}$ at a constant speed. Thus, the quantity $V_{AB}$
is not only measurable, but of practical use, as we will see later
on. But also of importance is the fact that the path-independence
of $V_{AB}$ implies the existence of a function $V\left(\mathbf{r}\right)$
such that 
\begin{equation}
V_{AB}\equiv V\left(\mathbf{r}_{B}\right)-V\left(\mathbf{r}_{A}\right).
\end{equation}
The function $V\left(\mathbf{r}\right)$ is called the \emph{electric
potential} of the field $\mathbf{E}$.

One important relationship between the electric field $\mathbf{E}$
and the potential $V$ is given by the gradient operator. The gradient
of $V\left(\mathbf{r}\right)$, denoted $\mbox{Grad}\left(V\left(\mathbf{r}\right)\right)$,
is defined, in Cartesian coordinates, as 
\begin{equation}
\mbox{Grad}\left(V\left(\mathbf{r}\right)\right)=\frac{\partial V\left(\mathbf{r}\right)}{\partial x}\hat{\bm{\textbf{\ensuremath{\imath}}}}+\frac{\partial V\left(\mathbf{r}\right)}{\partial y}\hat{\bm{\jmath}}+\frac{\partial V\left(\mathbf{r}\right)}{\partial z}\hat{\bm{\bm{k}}},\label{eq:gradient}
\end{equation}
where $\hat{\bm{\imath}}$, $\hat{\bm{\jmath}}$, and $\hat{\bm{k}}$
are the orthonormal basis vectors. The gradient $\mbox{Grad}\left(V\left(\mathbf{r}\right)\right)$
is the spatial derivative of $V$ at $\mathbf{r}$, and it forms a
vector field with the following properties. For a unitary vector $\hat{\mathbf{u}},$
$\mbox{Grad}\left(V\left(\mathbf{r}'\right)\right)\cdot\hat{\mathbf{u}}$
gives the rate of change of the function $V$ at point $\mathbf{r}'$
in the direction $\mathbf{\hat{u}}$. Thus, $\mbox{Grad}\left(V\left(\mathbf{r}'\right)\right)$
at $\mathbf{r}'$ is a vector, pointing at the direction where the
function $V$ changes the most, whose magnitude is the rate of change.
In other words, the direction perpendicular to $\mbox{Grad}\left(V\left(\mathbf{r}'\right)\right)$
points at the direction of the \emph{isopotential lines}. From this,
it is possible to prove that the function $V\left(\mathbf{r}\right)$
relates to the field by the expression 
\begin{equation}
\mathbf{E}(\mathbf{r})=-\mathrm{Grad}(V(\mathbf{r})),\label{eq:gradient-of-phi}
\end{equation}
i.e., the electric field is the negative gradient of the electric
potential.

It is easy to see that $V\left(\mathbf{r}\right)$ is not uniquely
defined, as any other function $V'\left(\mathbf{r}\right)$ given
by 
\begin{equation}
V'\left(\mathbf{r}\right)\equiv V\left(\mathbf{r}\right)-V_{0},\label{eq:Vprime}
\end{equation}
where $V_{0}$ is an arbitrary constant, also gives the same differences
of potential $V_{AB}$ between positions $\mathbf{r}_{A}$ and \textbf{$\mathbf{r}_{B}$},
and therefore the same gradient (for instance, when we take the limit
of $\Delta\mathbf{r}=\mathbf{r}_{B}-\mathbf{r}_{A}$ very small).
As an example, let us consider the electric field from a point particle
with charge $q$ situated at position $\mathbf{r}_{0}$, which is
given by 
\[
\mathbf{E}\left(\mathbf{r}\right)=\frac{q}{4\pi\epsilon_{0}}\frac{\mathbf{r}-\mathbf{r}_{0}}{\left(\mathbf{r}-\mathbf{r}_{0}\right)^{3}}.
\]
It is easy to show that a potential function satisfying \eqref{eq:line-integral-field}
is given by 
\begin{equation}
V(\mathbf{r})=-\frac{q}{4\pi\epsilon_{0}}\frac{1}{\left|\mathbf{r}-\mathbf{r}_{0}\right|}.\label{eq:point-particle-potential}
\end{equation}
For another potential function $V'$ given by $V'(\mathbf{r})=-\frac{q}{4\pi\epsilon_{0}}\frac{1}{\left|\mathbf{r}-\mathbf{r}_{0}\right|}+V_{0},$
we see at once that its gradient is exactly the same as for $V\left(\mathbf{r}\right)$,
since any constant added to $V$ would disappear when the derivative
in \eqref{eq:gradient} is taken. However, the potential \eqref{eq:point-particle-potential}
is often given in textbooks, as it has the feature that the potential
is zero in infinity ($V\rightarrow0$ when $\left|\mathbf{r}\right|\rightarrow\infty$),
which corresponds to a reference in a point infinitely distant from
$\mathbf{r}_{0}$.

As we mentioned above, the potential difference has practical use.
For actual measurements, directly observing the electric field is
very difficult, but measuring the electric work on a test particle,
related to the line integral of \eqref{eq:point-particle-potential}
between two points $\mathbf{r}_{A}$ and $\mathbf{r}_{B}$ is not.
This is done by inserting two probes (the electrodes, in the case
of the EEG) at points $\mathbf{r}_{A}$ and $\mathbf{r}_{B}$, and
creating a parallel circuit with high impedance, such that minimal
disturbance is created in the original potential. The current appearing
in this circuit is proportional to the potential difference $V(\mathbf{r}_{B})-V(\mathbf{r}_{A})$.
Thus, current measurements between different points can be used in
this way to construct a map of the potential function at different
positions (minus a reference value $V_{0}$, which is often arbitrarily
chosen such that $V(\mathbf{r}_{\mathcal{O}})=0$ at some reference
point $\mathcal{O}$). 

In a volume conductor, such as the brain, the electric field is a
useful quantity to know, as it is related to a current density $\mathbf{j}$
given by 
\begin{equation}
\mathbf{E}(\mathbf{r})=\rho\,\mathbf{j}(\mathbf{r}),\label{eq:relation-field-current}
\end{equation}
where $\rho=1/\sigma$ is the resistivity, given by the inverse of
the conductivity $\sigma$%
\footnote{Here, for simplicity, we are ignoring the fact that $\rho$ may depend
of both on position and time $\rho=\rho\left(\mathbf{r},t\right)$.
Furthermore, when the media is anisotropic, $\mathbf{E}\left(\mathbf{r}\right)$
is not necessarily in the same direction as $\mathbf{j}\left(\mathbf{r}\right)$,
and $\rho$ needs to be a rank-two tensor. However, at the scalp,
equation \eqref{eq:relation-field-current} is an excellent approximation
to EEG, as the $\rho$ is approximately isotropic in the tangential
direction \citep{rush1968current,nicholson_theory_1975,wolters2006influence,petrov2012anisotropic}. %
}. Equation \eqref{eq:relation-field-current} is sometimes referred
to as the vector form of Ohm's Law. In a conductive media with resistivity
$\rho$, a map of the potential function may be used to compute the
electric field $\mathbf{E}\left(\mathbf{r}\right)$ and, via equation
\eqref{eq:relation-field-current}, estimate current sources. Here
we must emphasize that electric potentials can \emph{never} be directly
measured; only potential differences between two points can.

Though in this paper we are interested in the surface Laplacian, for
completeness we should also mention the physical meaning of the Laplacian.
Such meaning comes from Gauss's Law, which in its differential form
given by Maxwell is written as 
\begin{equation}
\mathrm{Div}(\mathbf{E})=4\pi\rho_{Q},\label{eq:GaussLaw}
\end{equation}
where $\rho_{Q}$ is the charge density%
\footnote{Usually in physics the charge density is represented by $\rho$, but
here we use the subscript $Q$ to distinguish it from the resistivity,
defined as $\rho=1/\sigma$. %
}. Here $\mathrm{Div}\left(\mathbf{E}\right)$ is the divergence of
the electric field $\mathbf{E}$. The divergence is a linear differential
operator acting on $\mathbf{E}$, which in Cartesian coordinates,
where the electric field is represented by $\mathbf{E}=E_{x}\hat{\bm{\imath}}+E_{y}\hat{\bm{\jmath}}+E_{z}\hat{\bm{k}}$,
takes the form 
\[
\mbox{Div}\left(\mathbf{E}\right)=\frac{\partial E_{x}}{\partial x}+\frac{\partial E_{y}}{\partial y}+\frac{\partial E_{z}}{\partial z}.
\]
The divergence of $\mathbf{E}$ can be interpreted as a local measure
of the difference between how much field (technically, its flux) gets
into an infinitesimal volume surrounding the point where the divergence
is computed and how much of this field gets out. If the divergence
is zero, the same amount of field that gets into the infinitesimal
volume also gets out; if the divergence is negative, the amount of
field getting in is more than getting out (sink); and if the divergence
is positive, more field gets out than comes into the infinitesimal
volume (source). For this reason, the divergence of $\mathbf{E}$
is a measure of sources and sinks of the field. Thus, Gauss's law
has the immediate physical interpretation that electric charges are
sources of an electric field: if there are no electric charges, the
divergence of the field is zero. Now, from equation \eqref{eq:gradient-of-phi}
we have at once that 
\begin{equation}
\mathrm{Div}(\mathbf{E})=-\mathrm{Div}(\mathrm{Grad}(V)).\label{eq:E-Lap-relation}
\end{equation}
The divergence of the gradient, $\mathrm{Div}(\mathrm{Grad}(V))$,
is defined as the Laplacian%
\footnote{Notice that, since both $\mbox{Div}$ and $\mbox{Grad}$ are spatial
derivative operators, it follows that the Laplacian is a second-order
spatial derivative operator, since it is defined as the divergence
of the gradient of $V$. %
} of $V$, denoted by $\mbox{Lap}\left(V\right)\equiv\mathrm{Div}(\mathrm{Grad}(V))$.
Because it is a second spatial derivative, the Laplacian does not
have an interpretation that is as straightforward as the ones for
the divergence or the gradient. Its meaning is related to the mean
value of the potential around the point where it is computed. For
example, imagine that $\mbox{Lap}\left(V\right)$ is positive at some
point $\mathbf{r}'$ where $V$ has a local minimum (first derivative
is zero); such positive value means that for the nearby points surrounding
$\mathbf{r}'$, most of the values of $V$ are greater than $V\left(\mathbf{r}'\right)$.
So, $\mbox{Lap}\left(V\right)$ measures the mean value around the
point. Be that as it may, we can also think of the Laplacian of $V$
as simply the divergence of $\mathbf{E}$. Thus, from \eqref{eq:E-Lap-relation}
and \eqref{eq:GaussLaw} it follows that the Laplacian of the potential
is proportional to the electric field sources, i.e., 
\begin{equation}
\mbox{Lap}\left(V\right)=-4\pi\rho_{Q}.\label{eq:GaussLawLaplacian}
\end{equation}

The above discussion was intended to show the connection between the
field $\mathbf{E}$ to the scalar potential $V$. However, in its
generality, it does not take into account some specific characteristics
of the EEG. First, the EEG is measured over the scalp, which is geometrically
a curved surface. Second, such measurements are not directly of the
field, but are instead differences of potential associated to small
currents flowing in the closed circuit formed between the two electrodes
(often between a measuring and a reference electrode), the measuring
device, and the current flow density in the head. In the head, such
currents define a vector field consisting of currents from the brain,
dura surface, bone, and scalp. This current density is proportional
at each point to the electric field. But outside of the head, the
EEG measuring apparatus closes the circuit and measures the small
current associated with such system (despite how high the impedance
is) \citep{MettingvanRijn1990,MettingvanRijn1991}%
\footnote{This is the case for standard EEG systems, but not true for capacitive
measurements \citep{spinelli2010insulating}.%
}. Finally, because such measurements are on the scalp, they happen
where there is a significant change in conductivity (interface scalp/air).
All of those points have some consequence for the interpretations
of the field and Laplacian.

Starting with the Laplacian, the most common assumption is that sources
of interest to the EEG are inside the skull, and that there are no
sources in the scalp itself \citep{nunez_electric_2006}%
\footnote{This, of course, is an approximation, given that even EEG electrodes
can chemically generate currents. The discussion currents generated
by the scalp-electrode interface are beyond the scope of this paper,
and the readers are referred to \citep{MettingvanRijn1990,MettingvanRijn1991,huigen2002investigation,chi2010dry}%
}. But if we start with this assumption, we have that, from \eqref{eq:GaussLaw},
\begin{equation}
\mathrm{Lap}(V)=0.\label{eq:lap-0}
\end{equation}
 Using the expression for the Laplacian in Cartesian coordinates,
we have%
\footnote{Here the attentive reader may argue that \eqref{eq:lap-0} may not
be well-defined on the surface of the scalp because of its discontinuous
boundary conditions. First of all, it is important to point out that
the Laplacian is related through the right hand side of \eqref{eq:lap-0}
to a physically measurable quantity, and therefore always defined
at a point. Second, the ill-defined character of the Laplacian comes
from oversimplified mathematical models of the boundary conditions,
and can be eliminated by simply taking the lateral derivative at the
interface. For instance, for the equations below, such as \eqref{eq:cartesian-laplacian-zero},
if we think about the direction $\hat{\mathbf{z}}$ as perpendicular
to the scalp, with increasing values as we leave the scalp, then we
can think of the term involving a derivative with respect to $z$
as defined as the left derivative (outward bound). Finally, from a
practical point of view, such points where the derivatives are not
well defined consist of a set of measure zero, and therefore of little
practical importance. We refer the reader to \citet{Carvalhaes2014}
for details. %
} 
\begin{equation}
\frac{\partial^{2}V}{\partial x^{2}}+\frac{\partial^{2}V}{\partial y^{2}}+\frac{\partial^{2}V}{\partial z^{2}}=0.\label{eq:cartesian-laplacian-zero}
\end{equation}
Let us now choose the coordinate system such that the scalp is on
the plane $x,y$ (for small areas, this is a good approximation).
Then we can rewrite \eqref{eq:cartesian-laplacian-zero} as 
\begin{equation}
\frac{\partial^{2}V}{\partial x^{2}}+\frac{\partial^{2}V}{\partial y^{2}}=\frac{\partial E}{\partial z},
\end{equation}
where on the right hand side we used the relationship between the
field and the potential. Remembering that in a conductor $\mathbf{E}=\rho\mathbf{j}$,
it follows that 
\begin{equation}
\frac{\partial^{2}V}{\partial x^{2}}+\frac{\partial^{2}V}{\partial y^{2}}=\rho\frac{\partial j_{z}}{\partial z}.\label{eq:surface-laplacian-current}
\end{equation}
The left hand side of \eqref{eq:surface-laplacian-current} is defined
as the \emph{surface Laplacian} of $V$, 
\begin{equation}
\mathrm{Lap}_{S}(V)=\frac{\partial^{2}V}{\partial x^{2}}+\frac{\partial^{2}V}{\partial y^{2}},\label{eq:surface_Laplacian_in_R2}
\end{equation}
and \eqref{eq:surface-laplacian-current} shows that $\mathrm{Lap}_{S}(V)$
is related to \emph{how abruptly the normal component of the current
changes in the direction perpendicular to the surface}. Thus, if $\mathrm{Lap}_{S}(V)$
is nonzero, conceptually, in the absence of sources, the current change
in this direction means a fanning out of the current lines, i.e.,
a spreading out of the currents. In the case of the scalp, this means
that a nonzero $\mathrm{Lap}_{S}(V)$ corresponds to \emph{diverging}
(with respect to the radial direction) \emph{current lines under the
scalp}, which is associated to the presence of a source of currents
inside the skull. Thus, we see that the surface Laplacian has a direct
relationship to currents under the scalp, and it can be shown that
it is related to electrical activities on the dura surface \citep{nunez_electric_2006}.

\subsection{Properties of the surface Laplacian\label{sub:Properties-of-Laplacian}}

\begin{figure}
\noindent \centering{}\includegraphics[scale=0.35]{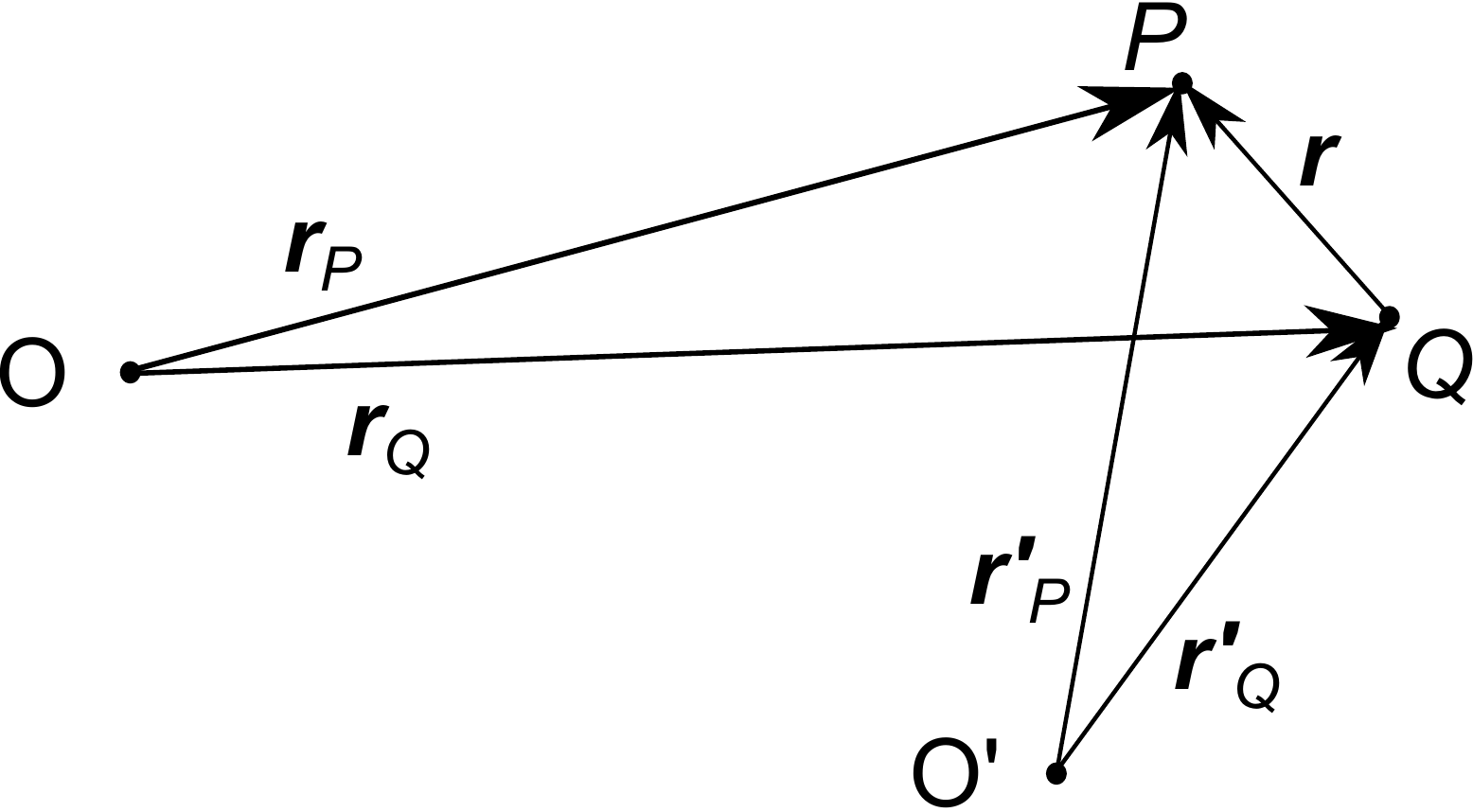}\protect\caption{\label{fig:Events}Events $P$ and $Q$ as observed from $\mathcal{O}$
and $\mathcal{O}'$. }
\end{figure}

Equipped with an interpretation of the surface Laplacian in terms
of currents and fields, we now turn to some its important properties.
Let us start with one of the main characteristics of the surface Laplacian:
it is \emph{reference free}. This may seem puzzling to some, as in
most cases the surface Laplacian is computed from the scalp potential,
which is itself a reference-dependent quantity. To understand how
this is possible, it is worth examining a simple example from geometry.
Imagine we have two points, $P$ and $Q$, representing the locations
of two events of interest. From an observer using the coordinate system
${\cal O}$ (see Figure \ref{fig:Events}), the positions of $P$
and $Q$ are given by the (reference-dependent) vectors $\mathbf{r}_{P}$
and $\mathbf{r}_{Q}$. That the position vectors of $P$ and $Q$
are reference dependent can be seen by the simple fact that another
observer using coordinates ${\cal O}'$ describe such positions by
the vectors $\mathbf{r}'_{P}$ and $\mathbf{r}'_{Q}$, which are clearly
different from $\mathbf{r}_{P}$ and $\mathbf{r}_{Q}$. Thus, the
geometrical meaning of, say, the length of the position vector $\mathbf{r}_{P}$,
is not related to the event of interest only, but to a combination
of such event and an \emph{arbitrary} choice of reference system,
making this quantity reference dependent.

However, there are many geometrically reference-free quantities, i.e.,
quantities that look the same for ${\cal O}$ and ${\cal O}'$, that
can be constructed from the reference-dependent positions $\mathbf{r}_{P}$
and $\mathbf{r}_{Q}$. For example, the vector connecting $P$ and
$Q$ is reference-free, since it is given by $\mathbf{r}=\mathbf{r}_{P}-\mathbf{r}_{Q}=\mathbf{r}'_{P}-\mathbf{r}'_{Q}$.
Notice that $\mathbf{r}$ has the feature of depending only on characteristics
of the system of interest, $P$ and $Q$. Another important reference-free
quantity is the distance $d$ between $P$ and $Q$, defined by $d^{2}=\mathbf{r}^{2}=\left(\mathbf{r}_{P}-\mathbf{r}_{Q}\right)\cdot\left(\mathbf{r}_{P}-\mathbf{r}_{Q}\right)$.
Thus, from reference-dependent geometrical quantities it is possible
to obtain reference-free ones. 

For the EEG, the reference-free property is not unrelated to the fact
that the surface Laplacian has physical meaning, whereas the scalp
potential does not%
\footnote{Furthermore, this property is connected to claims that the EEG topography
is not affected by changes in the reference electrode, which corresponds
to the statement that differences of potential between two points
across the topography are invariant with respect to the reference. %
}. For instance, as we saw above, the electric field is operationally
defined by its effect on a test charge. In a medium where the electric
conductivity $\sigma$ is non-zero, such as in the brain, skull, or
scalp, this effect translates into a (local) current given by 
\begin{equation}
\mathbf{E}=\rho\thinspace\mathbf{j},
\end{equation}
where $\rho=1/\sigma$ is the resistivity and $\mathbf{j}$ is the
electric current density. The electric potential, on the other hand,
is not operationally defined; instead, the \emph{difference} of potential
between two points is. For instance, as we saw above, in most EEG
experiments what is measured is actually the current flowing between
two electrodes, which is proportional to the difference of potential
between the two electrode positions. In practice, as mentioned before,
potential differences between two electrodes are estimated by the
insertion of a circuit in parallel with them, such that currents in
this circuit can be measured \citep{nunez_electric_2006} as a proxy
for the difference of potential; the effect of such measurement is
minimized by designing the added circuit to have high impedance \citep{MettingvanRijn1990,MettingvanRijn1991}.
So, if we define a potential function $V\left(\mathbf{r}\right)$,
the value of such function tells us nothing about measurable quantities
in $\mathbf{r}$; what is measurable is the potential difference between
two observation points, $\Delta V=V\left(\mathbf{r}'\right)-V\left(\mathbf{r}\right)$.
In other words, from a physical point of view it makes no sense to
say that the potential at a point $P$ is, for instance, $V\left(\mathbf{r}_{P}\right)$;
however, it does make sense to say that the electric field at point
$P$ is $\mathbf{E}\left(\mathbf{r}_{P}\right)$. From the electric
field we can obtain the Laplacian, given by $\mathrm{Div}(\mathbf{E})$.
Thus, the Laplacian is simply the divergence of $\mathbf{E}$, and
as such is nonzero where there are sinks or sources of electric field.

In practice, directly measuring the scalp field or surface Laplacian
is technically very difficult. However, the usual measurements of
differences of potential on the scalp is technically easy. Because
of this, it is common, as we mentioned above, to use the observed
differences of potential (often defined against a reference electrode)
to calculate the surface Laplacian or the electric field (see Section
\ref{sec:Computational-Aspects/Methods} for computational methods).

Of course, being reference free does not mean that the surface Laplacian
of $V$ computed from an EEG is not subject to problems due to a bad
choice of reference electrodes. To understand this, let us imagine
that we have two nearby points of interest on the scalp, say $P_{1}$
and $P_{2}$. Their electric potential difference $\Delta V_{12}=V\left(\mathbf{r}_{2}\right)-V\left(\mathbf{r}_{1}\right)$,
where $\mathbf{r}_{1}$ and $\mathbf{r}_{2}$ are the positions of
$P_{1}$ and $P_{2}$, is physically relevant (and reference free),
since it is proportional to the work done by the electric field on
a test charge moving from $P_{1}$ to $P_{2}$. This potential difference
can be observed either by measuring the current flow between $P_{1}$
and $P_{2}$, which is proportional to $\Delta V_{12}$, or by measuring
the current flow between each point $P_{1}$ and $P_{2}$ with respect
to another point $R$, the place of a reference electrode. Since $\Delta V_{1R}=V\left(\mathbf{r}_{1}\right)-V\left(\mathbf{r}_{R}\right)$
and $\Delta V_{2R}=V\left(\mathbf{r}_{2}\right)-V\left(\mathbf{r}_{R}\right)$,
it follows that 
\begin{equation}
\Delta V_{12}=\Delta V_{2R}-\Delta V_{1R}.\label{eq:differences}
\end{equation}
As we will see below in Section \ref{sec:Computational-Aspects/Methods},
both the electric field and the Laplacian may be computed from terms
such as $\Delta V_{2R}-\Delta V_{1R}$. 

Expressions like \eqref{eq:differences} overlook an important problem:
recording and data acquisition conditions. The choice of a reference
faces two important factors. First, a choice of a reference electrode
that is too far away (say, on the left foot) can cause a reduction
on the measured current between, say, $P_{1}$ and $R$, due to the
sources of interest in the skull. For instance, for usual multipole
sources in the skull, the potential decreases very rapidly as we move
away from it, becoming constant at small distances \citep{jackson_classical_1999}.
However, the resistance between $P_{1}$ and $R$ keeps increasing
as we move $R$ further away. Thus, the effectively measured current
in the circuit decreases, and such a decreased current will translate
into a lower signal-to-noise ratio (an extreme case, discussed by
\citet{nunez_electric_2006} is when the electrode is place on a laboratory
wall, and therefore no current can flow, as there is no closed circuit).
Because of the presence of random noise, $\Delta V_{2R}-\Delta V_{1R}$
is not exactly the same as $\Delta V_{12}$ (neither is the actual
bipolar measurement), but a bad choice of electrode makes this problem
even more pronounced. But distance between electrodes is not the only
problem. Even a close electrode may be in a region where undesirable
potential generators are present (e.g., eye movement). The use of
such electrode locations as reference further aggravates the signal-to-noise
problem by adding extraneous physiological noise. 

The above discussion holds for the general case, being valid for all
cases where a quasi-static electric field interaction is involved
\citep{Haemaelaeinen1993}. We included such discussion here because
it is seldom present in the EEG review literature, but it bears relevance
to many of the concepts often overlooked. We now turn to a description
of the computational methods.

\section{Computational Methods\label{sec:Computational-Aspects/Methods}}

Although the arguments supporting the Laplacian technique are very
solid, it turns out that EEG experiments are ordinarily designed to
record the electric potential, and any information about its spatial
derivatives, including the surface Laplacian, needs to be estimated
from the recordings through a numerical procedure. There are several
methods available for surface Laplacian estimates, but some may be
preferable to others for particular purposes. The first Laplacian
estimates in the literature were performed by \citealp{hjorth_-line_1975}%
\footnote{We also acknowledge the work of \citet{Freeman1975}, who employed
a discretization scheme to approximate the Laplacian of extra-cellular
potentials in CSD analysis.%
}, who presented an approximation to equation~\eqref{eq:surface_Laplacian_in_R2}
based on \emph{finite difference}. This method is conceptually simple
and easy to implement, and for this reason still very popular. It
also does not suffer from some numerical problems that affect its
main alternatives, but has the downside of relying on assumptions
that are quite questionable empirically. In a further development,\emph{
mesh-free} methods \citep{perrin_mapping_1987,perrin_scalp_1987,perrin_spherical_1989}
were introduced, eliminating problems encountered in finite difference,
but other issues emerged, some of which still need attention. Our
goal here is to review both types of approaches and emphasize issues
related to each.

Two other topics of great importance are discussed in detail. The
first is the concept of the \emph{surface Laplacian matrix}, which
provides a means to significantly reduce the computational cost of
estimating the surface Laplacian. The discrepancydifference between
the computation cost of this approach and direct computation is overwhelmingvery
large and should not be overlooked, mainly when using mesh-free methods.
The second topic is related to the procedure of \emph{regularization}
to increase the quality of estimates. Besides reviewing this subject
and presenting a practical example with real data, we also suggest
a solution to an important problem arising in the regularization of
multiple frames of an EEG signal.

\subsection{Finite difference methods\label{sub:Finite-Differences}}

\begin{figure}
\noindent \centering{}\includegraphics[width=0.65\textwidth]{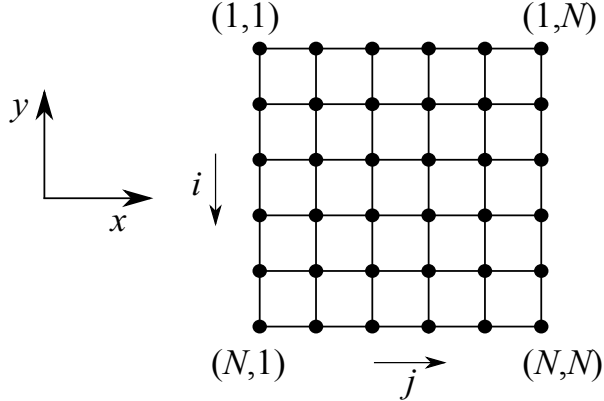}\protect\caption{Mapping from the continuous variables $x$ and $y$ into the grid
variables $i$ and $j$.\label{fig:xy_to_ij}}
\end{figure}

Generally speaking, finite difference is a \emph{discretization procedure}
in which a continuous domain is transformed into a mesh of discrete
points, where differential operations such as the Laplacian becomes
more manageable. Figure\,\ref{fig:xy_to_ij} depicts the construction
that will be used here. In this construction, each electrode occupies
an individual node of a regular grid specified by the discrete variables
$i$ and $j$, in replacement to the continuous variables $x$ and
$y$ used before. Two major assumptions are taking placemade. First,
the scalp surface is locallyapproximately flat, and second the measuring
electrodes are equidistant, forming a square grid of size $h$. The
first assumption leads to the following differential form in Cartesian
coordinates (see also Section \ref{sub:Physical-Interpretation})
\begin{equation}
\mathrm{Lap}_{s}(V)=\frac{\partial^{2}V}{\partial x^{2}}+\frac{\partial^{2}V}{\partial y^{2}}.\label{eq:two-dimensional-Laplacian}
\end{equation}
With the second assumption we are able to approximate $\mathrm{Lap}_{s}(V)$
at a central node $(i,j)$ by \citep[25.3.30]{Abramowitz1964} 
\begin{equation}
\mathrm{Lap}_{s}(V)\big|_{(i,j)}\approx\frac{V_{(i-1,j)}+V_{(i+1,j)}+V_{(i,j-1)}+V_{(i,j+1)}-4V_{(i,j)}}{h^{2}}.\label{eq:Hjorth-approximation}
\end{equation}
The mathematics behind this approximation is nontrivial and for completeness
we present a detailed explanation in Appendix A. 

As expected from the discussion of Section \ref{sub:Properties-of-Laplacian},
this approximation is reference-free. To see this, simply replace
$V$ with $V^{\prime}-V_{\mathrm{ref}}$ at each node (including the
central node) and note that the common reference $V_{\mathrm{ref}}$
is canceled. The right hand side of \eqref{eq:Hjorth-approximation}
can be interpreted as a change in reference to the average over the
four nearest neighbors of node $(i,j)$.%
\footnote{Notice that this reference change happens locally, i.e. it is different
for each electrode position, which means that the Laplacian is not
simply another reference scheme. %
} If the reference electrode is located on the scalp we can reconstruct
the Laplacian at this location, where $V=0$, by computing 
\begin{equation}
\mathrm{Lap}_{s}(V_{\mathrm{ref}})\approx\frac{V_{(i-1,j)}+V_{(i+1,j)}+V_{(i,j-1)}+V_{(i,j+1)}}{h^{2}}.\label{eq:Hjorth-approximation-reference}
\end{equation}
The multiplicative factor $1/h^{2}$ in \eqref{eq:Hjorth-approximation}
and \eqref{eq:Hjorth-approximation-reference} ensures the correct
physical unit of $\mathrm{Lap}_{s}(V)$, which is Volt per centimeter
square (V/cm$^{2}$), but in many situations this is ignored, as originally
done in \citet{hjorth_-line_1975}.

\begin{figure}
\noindent \centering{}\includegraphics[width=1\textwidth]{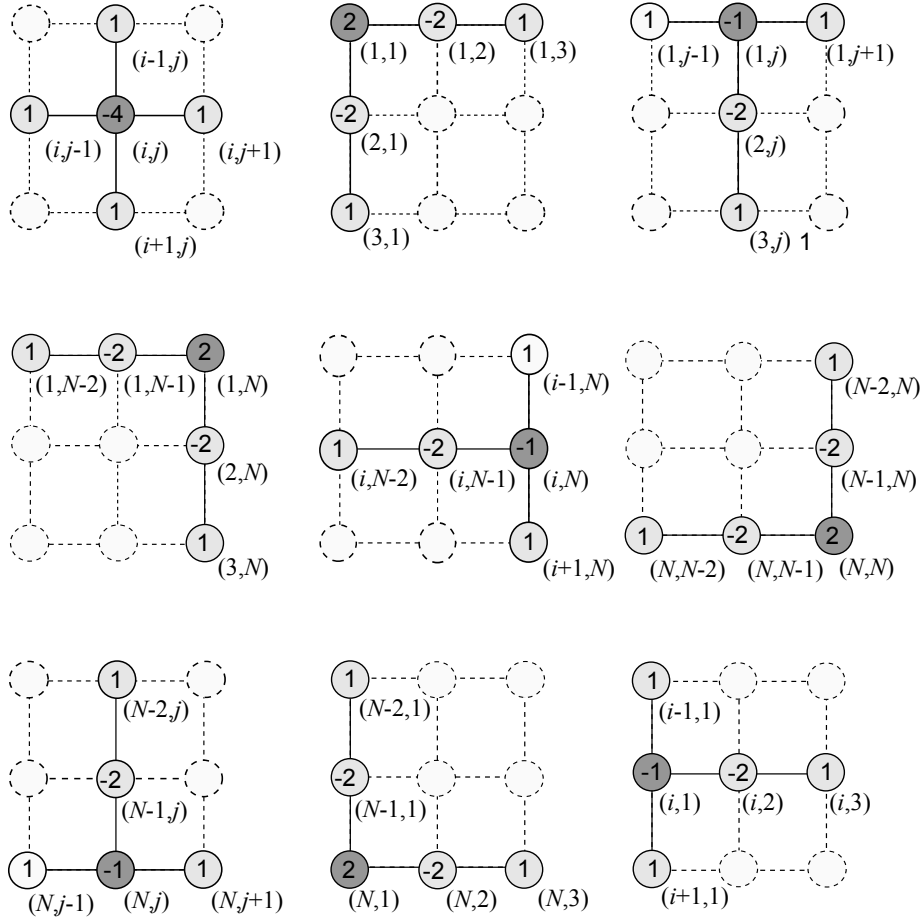}\protect\caption{Geometric arrangements and weights to approximate
$\mathrm{Lap}_{s}(V)$ at different locations on a grid.}\label{fig:Five-point-stencils}
\end{figure}

Approximation \eqref{eq:Hjorth-approximation} is usually refer to
as \emph{Hjorth's approximation}. A limitation of this approximation
is that it applies only to electrodes located at a central node, thus
not accounting for estimates along the border of the electrode grid.
For border electrodes, Hjorth suggested a less accurate approximation
using just three electrodes aligned along the border. In an effort
to account for estimations at all electrode sites, we developed the
scheme of Figure\,\ref{fig:Five-point-stencils}. For instance, estimates
along the left border, where $j=1$, are given by 
\begin{equation}
\mathrm{Lap}_{s}(V_{(i,1)})\approx\frac{V_{(i-1,1)}+V_{(i+1,1)}-2V_{(i,2)}+V_{(i,3)}-V_{(i,1)}}{h^{2}},\label{eq:Hjorth-left-border}
\end{equation}
and at the upper-left corner 
\begin{equation}
\mathrm{Lap}_{s}(V_{(1,1)})\approx\frac{-2V_{(1,2)}+V_{(1,3)}-2V_{(2,1)}+V_{(3,1)}+2V_{(1,1)}}{h^{2}}.\label{eq:Hjorth-upper-left-corner}
\end{equation}
Our calculations are shown in Appendix~A. As it occurs with approximation
\eqref{eq:Hjorth-approximation}, these expressions average the potential
with weights that sum to zero, thus canceling the reference potential.

The procedure described in Appendix~A can be used to obtain approximations
combining more than five electrodes. However, increasing complexity
does not result necessarily into more accurate estimates. Approximations
for unevenly-spaced electrodes can also be derived, but we will not
address this in this paper \citep[cf.][]{Tenke1998}\textbf{.}

Finite difference provides us with a simple measure of accuracy in
terms of powers of the discretization parameter $h$. As explained
in Appendix~A, approximation\emph{ }\eqref{eq:Hjorth-approximation}
is said to be accurate to the second-order of $h$, shorthand ${\cal O}(h^{2})$,
while for peripheral electrodes the five-point approximations \eqref{eq:Hjorth-left-border}
and \eqref{eq:Hjorth-upper-left-corner} are accurate to ${\cal O}(h)$.
The higher the order, the more accurate is the approximation \citep{Abramowitz1964}.
Therefore, in principle, making $h$ small by using high-density electrode
arrays should reduce discretization error and improve estimates. Moreover,
a small $h$ is useful also to reduce error due to the unrealistic
geometry of the plane scalp model, which is more prominent when averaging
over sparse electrode sites. In the absence of an appropriate mechanism
for regulation, high-density arrays can also help reduce influence
of non-local activities from unrelated sources. But exceptions are
expected to occur. In fact, all this reasoning was based on the assumption
of a point electrode, which ignores boundary effects produced by an
actual electrode of finite size. Perhaps for this reason, and due
to factors that we are not able to point out, \citet{McFarland1997}
reported better estimates using next-nearest-neighbor approximations
in comparison to nearest-neighbor approximations. Similar conclusion
was drawn by \citet{Mueller-Gerking1999} in the context of EEG classification.
In fact, next-nearest neighbor approximations have been adopted in
CSD analysis since \citet{Freeman1975} as a way to reduce high-frequency
noise in the vicinity of an electrode. The trade-off between noise
attenuation and loss of spatial resolution in low-density arrays was
discussed by \citet{Tenke1993}. In fact, lower-density estimates
has proven to be quite useful more than once, as described by \citet{Kayser2006}
in the context of ERP analysis. Finally, we point out that evidence
for electrode bridges in high-density EEG recordings was identified
as a problem caused typically by electrolyte spreading between nearby
electrodes \citep{Tenke2001,Greischar2004,Alschuler2014}.

\subsection{Smoothing thin-plate spline Laplacians\label{sub:Thin-Plate-Splines}}

As an alternative to finite difference, \emph{mesh-free} (or \emph{grid-free})
methods allow for a much more flexible configuration of electrodes
and are not restricted to the planar scalp model. In this case, the
Laplacian differentiation is performed analytically on a continuous
function built from the data. Two ways of obtaining this function
are through interpolation or a parametrization procedure known as
\emph{data smoothing}. In the context of interpolation, the problem
to be solved can be formulated as follows: given a set of measurements,
say a potential distribution $V_{1},\cdots,V_{N}$ recorded from locations
$\mathbf{r}_{1},\cdots,\mathbf{r}_{N}$ (the electrode locations),
find a real function $f(\mathbf{r})$ such that $f(\mathbf{r}_{i})=V_{i}$
for $i=1,\cdots,N$. For data smoothing, this constraint is replaced
with the condition that $f(\mathbf{r}_{i})$ should fit the data closely,
but not exactly to reduce variance due to noise. The method of \emph{thin-plate
splines} discussed here is a prime tool for both cases.

Formally, the spline function is the \emph{unique} solution to the
problem of finding a function $f$ that minimizes \citep[p.140]{duchon_splines_1977,meinguet_1979,wahba_spline_1990,Hastie2009}
\begin{equation}
\mathrm{RSS}(f,\lambda)=\frac{1}{N}\sum_{i}\left(V_{i}-f(\mathbf{r}_{i})\right)^{2}+\lambda\thinspace J_{m}\left[f(\mathbf{r})\right],\label{eq:TPS_Variational_Problem}
\end{equation}
where it was assumed that $\mathbf{r}_{1}\neq\mathbf{r}_{2}\neq\cdots\neq\mathbf{r}_{N}$
(which is always true for actual electrodes) and $J_{m}\left[f\right]$
is a measure of roughness $f$ in terms of its $m$-th-order partial
derivatives (the reader is referred to \citealp{wahba_spline_1990}
for details). The parameter $\lambda$ is called the \emph{regularization
parameter.} This parameter trades off between goodness of fit ($\lambda=0$)
and smoothness ($\lambda>0$). In other words, if $\lambda=0$ then
the interpolation condition $f(\mathbf{r}_{i})=V_{i}$ is satisfied,
whereas for $\lambda>0$ the spline function is smooth and violates
the interpolation condition. In three-dimensional Euclidean space
the solution to this problem has the form \citep{wahba_spline_1990,Wood2003}

\begin{equation}
f_{\lambda}(\mathbf{r})=\sum_{i=1}^{N}c_{i}\left|\mathbf{r}-\mathbf{r}_{i}\right|^{2m-3}+\sum_{\ell=1}^{M}d_{\ell}\nu_{\ell}(\mathbf{r}),\quad(m\ge2).\label{eq:TPS_Interpolating_Function}
\end{equation}
The parameters $m$ and $M$ are integer numbers satisfying $M=\binom{m+2}{3}$
and $M<N$. In addition, $m$ must be greater than 2 for the surface
Laplacian differentiation to be well-defined. The functions $\nu_{1},\cdots,\nu_{M}$
are linearly independent polynomials in three variables of degree
less than $m$ (Tthe reader is referred to \citet{Carvalhaes2011}
for the details of how to generate these polynomials). 

Using matrix notation we can express the unknowns $c_{i}$'s and $d_{i}$'s
as the solution to the linear system \citep{duchon_splines_1977,meinguet_1979,wahba_spline_1990,Green1994,Eubank:1999jk}
\begin{equation}
\left(\begin{array}{cc}
\mathbf{K}+N\lambda\mathbf{I} & \mathbf{T}\\
\mathbf{T}^{T} & 0
\end{array}\right)\left(\begin{array}{c}
\mathbf{c}\\
\mathbf{d}
\end{array}\right)=\left(\begin{array}{c}
\mathbf{v}\\
\mathbf{0}
\end{array}\right),\label{eq:TPS_Linear_System}
\end{equation}
where $\mathbf{c}=(c_{1},\cdots,c_{N})^{T}$, $\mathbf{d}=(d_{1},\cdots,d_{M})^{T}$,
$(\mathbf{K})_{ij}=\left|\mathbf{r}_{j}-\mathbf{r}_{i}\right|^{2m-3}$,
$(\mathbf{T})_{ij}=\nu_{j}(\mathbf{r}_{i})$, and $\mathbf{v}=(V_{1},\cdots,V_{N})^{T}$
is the given potential distribution. The superscript $T$ indicates
transpose operation and $\mathbf{I}$ is the $N\times N$ identity
matrix. This system has the formal solution \citep{wahba_spline_1990}\begin{subequations}\label{eq:TPS_Solution_for_c_and_d}
\begin{gather}
\mathbf{c}=\mathbf{Q}_{2}\left[\mathbf{Q}_{2}^{T}\left(\mathbf{K}+N\lambda\mathbf{I}\right)\mathbf{Q}_{2}\right]^{-1}\mathbf{Q}_{2}^{T}\mathbf{v},\label{eq:TPS_Solution_for_c}\\
\mathbf{R}\mathbf{d}=\mathbf{Q}_{1}^{T}\left(\mathbf{v}-\mathbf{K}\mathbf{c}-N\lambda\mathbf{c}\right),\label{eq:TPS_Solution_for_d}
\end{gather}
\end{subequations}where $\mathbf{Q}_{1}\in\mathbb{R}^{N\times N}$,
$\mathbf{Q}_{2}\in\mathbb{R}^{N\times(N-M)}$, and $\mathbf{R}\in\mathbb{R}^{N\times M}$
derive from the \emph{QR-factorization }of $\mathbf{T}$ %
\footnote{The Matlab command for the QR-factorization of $\mathbf{T}$ is \texttt{{[}Q,R{]}=qr(T)}.%
}: 
\begin{equation}
\mathbf{T}=[\mathbf{Q}_{1}\thinspace|\thinspace\mathbf{Q}_{2}]\begin{pmatrix}\mathbf{R}\\
 & \mathbf{0}
\end{pmatrix}.
\end{equation}
With the above definitions we can express the smoothed potential at
the electrode locations as 
\begin{equation}
\mathbf{v}_{\lambda}=\mathbf{K}\mathbf{c}+\mathbf{T}\mathbf{d}.
\end{equation}
By introducing the matrices $(\tilde{\mathbf{K}})_{ij}=\mathrm{Lap}\left(\left|\mathbf{r}_{j}-\mathbf{r}_{i}\right|^{2m-3}\right)$
and $(\tilde{\mathbf{T}})_{ij}=\mathrm{Lap}\left(\nu_{j}(\mathbf{r}_{i})\right)$,
we obtain an analogous expression for the Laplacian estimate, which
is
\begin{equation}
\mathrm{Lap}_{S}(\mathbf{v})=\tilde{\mathbf{K}}\mathbf{c}+\tilde{\mathbf{T}}\mathbf{d}.
\end{equation}

This approach has a major deficiency, which is that it does not work
if the data points $\mathbf{r}_{1},\cdots,\mathbf{r}_{N}$ are located
on a spherical or an ellipsoidal surface. To understand this limitation,
consider the case $m=3$, for which the matrix $\mathbf{T}$ can be
expressed as \citep{Carvalhaes2011} 
\begin{equation}
\mathbf{T}=\left(\begin{array}{cccccccccc}
1 & x_{1} & y_{1} & z_{1} & x_{1}^{2} & x_{1}y_{1} & x_{1}z_{1} & y_{1}^{2} & y_{1}z_{1} & z_{1}^{2}\\
\vdots & \vdots & \vdots & \vdots & \vdots & \vdots & \vdots & \vdots & \vdots & \vdots\\
1 & x_{N} & y_{N} & z_{N} & x_{N}^{2} & x_{N}y_{N} & x_{N}z_{N} & y_{N}^{2} & y_{N}z_{N} & z_{N}^{2}
\end{array}\right).\label{eq:TPS_T_Matrix}
\end{equation}
Since on spherical and ellipsoidal surfaces the coordinates $x_{i}$,
$y_{i}$, and $z_{i}$ are subject to%
\footnote{Note that $a=b=c$ on the sphere.%
} 
\begin{equation}
\frac{x_{i}^{2}}{a^{2}}+\frac{y_{i}^{2}}{b^{2}}+\frac{z_{i}^{2}}{c^{2}}=1,\quad a,b,c>0,\label{eq:TPS_Ellipsoid}
\end{equation}
the 1st, 5th, 8th, and 10th columns of $\mathbf{T}$ are \emph{linearly
dependent}%
\footnote{That is, summing these columns with weights $-1$, $1/a^{2}$, $1/b^{2}$
and $1/c^{2}$ yields a zero vector.%
}. Hence, we can always express one of these columns in terms of the
others. This is equivalent to say that $\mathbf{T}$ has 10 columns
but only 9 are linearly independent. This makes the linear system
\eqref{eq:TPS_Linear_System} \emph{singular}, so that the unknowns
$c_{i}$'s and $d_{i}$'s cannot be uniquely determined.

The singularity of \eqref{eq:TPS_Linear_System} on spheres and ellipsoids
affects only the transformation parametrized by $\mathbf{d}$, leaving
intact the transformation specified by $\mathbf{c}$. It is therefore
natural to try a minimum norm solution to this problem by determining
$\mathbf{d}$ using the \emph{pseudo-inverse} of $\mathbf{R}$, i.e.,
$\mathbf{d}=\mathbf{R}^{+}\mathbf{Q}_{1}^{\prime}\left(\mathbf{v}-\mathbf{K}\mathbf{c}-N\lambda\mathbf{c}\right)$.
This approach was proposed by \citet{Carvalhaes2011} (see also \citet{Carvalhaes2013})
and was evaluated in a problem involving the localization of cortical
activity on spherical and ellipsoidal scalp models. Simulations using
over 30,000 configurations of radial dipoles resulted in a success
rate above 94.5\% for the correct localization of cortical source
at the closest electrode. This rate improved to 99.5\% for the task
of locating the electrical source at one of the two closest electrode.
Prediction error occurred more often for sources generating very small
or very large peaks in the amplitude of the potential, but it decreased
substantially with increasing the number of electrodes in the simulation.
In the same study, but now using empirical data, the surface Laplacian
of \eqref{eq:TPS_Interpolating_Function} outperformed finite difference
and the method of spherical splines discussed below.

Similar to finite difference, splines and other methods discussed
below are usually affected by the density of the electrode array.
In general, high-density arrays result in more accurate estimates.
However, some preparation issues such as electrolyte bridges are mainly
a characteristic of high-density estimates and should be a concern,
as it is for finite difference \citep{Tenke2001}. Another concern
is that by increasing the number of electrodes the system \eqref{eq:TPS_Linear_System}
becomes increasingly sensitive to error (noise) in the input vector
$\mathbf{v}$. According to results in \citet{Sibson1991} it is expected
that montages with up to 256 electrodes result in reliable estimates
for double-precision calculations, but there is one more issue requiring
special attention.

It follows from \eqref{eq:TPS_Solution_for_c} that in order to obtain
the coefficient vector $\mathbf{c}$ the matrix $[\mathbf{Q}_{2}^{T}\left(\mathbf{K}+N\lambda\mathbf{I}\right)\mathbf{Q}_{2}]$
needs to be inverted. However, it turns out that this inverse operation
is accurate only if the condition number of $\left(\mathbf{K}+N\lambda\mathbf{I}\right)$
is relatively small.%
\footnote{The condition number of a matrix $\mathbf{A}$ is a real number that
estimates the loss of precision in inverting this matrix numerically.
This number, denoted by $\mathrm{cond}(\mathbf{A})$, is usually computed
as the ratio of the largest to smallest eigenvalue of $\mathbf{A}$,
implying that $\mathrm{cond}(\mathbf{A})\ge1$. The larger the value
of $\mathrm{cond}(\mathbf{A})$, the more inaccurate the inversion
operation.%
}. In principle, a condition number up to $10^{12}$ should be acceptable
for splines \citep{Sibson1991}, but this upper bound can easily be
exceeded as $\lambda$ decreases toward zero depending on the electrode
configuration. In fact, even for a small number of electrodes the
condition number of $\left(\mathbf{K}+N\lambda\mathbf{I}\right)$
can be so high that is impossible to invert $[\mathbf{Q}_{2}^{T}\left(\mathbf{K}+N\lambda\mathbf{I}\right)\mathbf{Q}_{2}]$
without regularization. This problem holds also for the algorithm
of spherical splines discussed in the next section, and \citet{Bortel2007}
discussed the same problem in the context of realistic Laplacian techniques.
In addition to compensating for conditioning issues, the regularization
parameter $\lambda$ determines also the ability of smoothing splines
to account for spatial noise in the data. This problem will be addressed
in Section\,\ref{sec:Data-Smoothing}.

\subsection{Smoothing spherical splines\label{sub:Spherical-Splines}}

For the particular case of data points on spheres, Wahba developed
a pseudo-spline method that circumvent the singularity of \eqref{eq:TPS_Linear_System}
by replacing the Euclidean distance with the geodesic distance. This
method, called \emph{spherical splines}, was used by \citet{perrin_spherical_1989}
to developed one of the most popular surface Laplacian methods in
the literature. In this case, the smoothing/interpolating function
is defined as \begin{subequations}\label{eq:SS_Interpolant} 
\begin{eqnarray}
f_{sph}(\mathbf{r}) & = & \sum_{i=1}^{N}c_{i}\, g_{m}(\mathbf{r},\mathbf{r}_{i})+d,\label{eq:SS_Interpolatiing_Function}
\end{eqnarray}
where the parameters $c_{1},\cdots,c_{N}$ and $d$ are determined
by \eqref{eq:TPS_Solution_for_c_and_d} and 
\begin{equation}
g_{m}(\mathbf{r},\mathbf{r}_{i})=\frac{1}{4\pi}\sum_{\ell=1}^{\infty}\frac{2\ell+1}{\ell^{m}(\ell+1)^{m}}P_{\ell}(\mathbf{\hat{r}}\cdot\mathbf{\hat{r}}_{i}).\label{eq:SS_Kernel}
\end{equation}
\end{subequations}The vectors $\mathbf{r}$ and $\mathbf{r}_{i}$'s
have length equal to the head radius and carets denote unit vectors,
i.e., $\hat{\mathbf{r}}=\mathbf{r}/r$ and $\hat{\mathbf{r}_{i}}=\mathbf{r}_{i}/r$,
so that the scalar product $\mathbf{\hat{r}}\cdot\mathbf{\hat{r}}_{i}$
gives the cosine of the angle between $\mathbf{r}$ and $\mathbf{r}_{i}$.
The parameter $m$ is an integer larger than 1. The univariable functions
$P_{\ell}$'s are Legendre polynomials of degree $\ell$. These polynomials
occur in many problems in mathematical physics and their properties
are listed in many places \citet[e.g.][]{Abramowitz1964}. The infinity
summation over Legendre polynomials generates progressively higher
spatial frequencies which are weighted by the factor $(2\ell+1)/\ell^{m}(\ell+1)^{m}$.
Since this factor decreases monotonically with $\ell$, it acts as
a Butterworth filter that downweights high-frequency spatial components
to produce a smooth result.

\begin{figure}
\noindent \centering{}\includegraphics[width=0.37\textwidth]{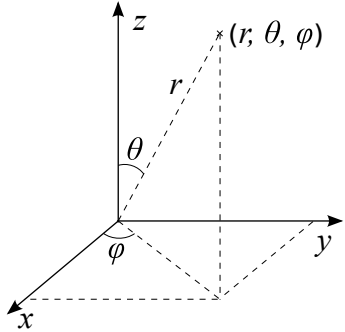}
\protect\caption{Spherical coordinate system.\label{fig:Spherical-coordinate-system.}}
\end{figure}

To apply the surface Laplacian to \eqref{eq:SS_Interpolant} we use
spherical coordinates. Our convention is that of Figure\,\ref{fig:Spherical-coordinate-system.}:
the polar angle $\theta\in[0,\pi]$ is measured down from the vertex,
and the azimuthal angle $\varphi\in[0,2\pi]$ increases counterclockwise
from the $x$ axis, which is directed towards the nasion. With this
convention, 
\begin{equation}
\mathrm{Lap}_{s}(f)=\frac{1}{r^{2}\,\sin\theta}\frac{\partial}{\partial\theta}\left[\sin\theta\frac{\partial f}{\partial\theta}\right]+\frac{1}{r^{2}\sin^{2}\theta}\frac{\partial^{2}f}{\partial\varphi^{2}}.\label{eq:SS_Formula_for_the_Laplacian}
\end{equation}
It can be proved that the Legendre polynomials satisfy \citep[p. 110]{jackson_classical_1999}
\begin{equation}
\mathrm{Lap}_{s}(P_{\ell}(\mathbf{\hat{r}}\cdot\mathbf{\hat{r}}_{i}))=-\frac{\ell(\ell+1)}{r^{2}}P_{\ell}(\mathbf{\hat{r}}\cdot\mathbf{\hat{r}}_{i}),\label{eq:SS_Laplacian_of_Legendre Polynomials}
\end{equation}
thus leading to 
\begin{equation}
\mathrm{Lap}_{s}(f(\mathbf{r}))=-\frac{1}{r^{2}}\sum_{i=1}^{N}c_{i}g_{m-1}(\mathbf{r},\mathbf{r}_{i}),\quad m>1,\label{eq:SS_Spherical Laplacian}
\end{equation}
which is known as the \emph{spherical surface Laplacian of $f(\mathbf{r})$.}

In comparison to the previous approach, the method of spherical splines
has the advantage of providing a simple expression for the surface
Laplacian of $f(\mathbf{r})$, although it is restricted to spherical
geometry only. Additionally, the geodesic distance ensures that the
system \eqref{eq:TPS_Linear_System} is non-singular, so that the
coefficient $d$ can be determined without using the pseudo-inverse
of $\mathbf{R}$. However, some caution is required. Equation \eqref{eq:SS_Spherical Laplacian}
is easy to be implemented, except for the fact that the term $g_{m-1}(\mathbf{r},\mathbf{r}_{i})$
corresponds to an infinite summation. This summation needs to be truncated
for evaluation, and truncation error should never be overlooked. The
terms that are left out in the truncation of $g_{m-1}(\mathbf{r},\mathbf{r}_{i})$
are Legendre polynomials of higher-degree (large $\ell$) that account
for high-frequency spatial features. Too much filtering of such features
can overcome the potential for improvement in spatial resolution.
In view of that, it is important to keep as many Legendre polynomials
as possible in the truncated series. Another point worth considering
is the factor $(2\ell+1)/\ell^{m}(\ell+1)^{m}$ that multiplies the
Legendre polynomials in equation~\eqref{eq:SS_Kernel}. This factor
falls off as $1/\ell^{2m-1}$. As $\ell$ increases, it approaches
zero rapidly, resulting in a quick cancellation of higher-degree polynomials.
The value assigned to $m$ must take this effect into account; the
larger the value of $m$, the more dramatic is the reduction in high-frequency
features. Typically, a value of $m$ between 2 and 6 provides satisfactory
results for simulation and data analysis \citep{babiloni_performances_1995,Barros2006,Carvalhaes2011}.

\subsection{Other approaches}

There are many other possible ways to estimate the surface Laplacian
of EEG signals. In this section we briefly review the methods proposed
by \citet{yao_high-resolution_2002} and Nunez and collaborators \citep{law_high-resolution_1993,nunez_electric_2006}.
Similar to thin-plate splines, Yao's approach uses interpolation based
on radial functions to estimate the surface Laplacian of the potential.
The interpolating function has the general form 
\begin{equation}
f_{\mathrm{RBF}}(\mathbf{r})=\sum_{i=1}^{N}c_{i}e^{-\frac{m}{\pi^{2}}S(\mathbf{r},\mathbf{r}_{i})}+b,\label{eq:Other_Approaches-Yao}
\end{equation}
where $S(\mathbf{r},\mathbf{r}_{j})$ measures the arc of circle connecting
$\mathbf{r}$ and $\mathbf{r}_{i}$ and the parameters $c_{i}$ and
$b$ are subject to 
\begin{equation}
f_{\mathrm{RBF}}(\mathbf{r}_{i})=V_{i},\quad i=1,\cdots,N,\label{eq:RBF-linear-system}
\end{equation}
where $\mathbf{r}_{i}$ is the location of the $i$th electrode and
$V_{i}$ is the value of the potential at that location. Because the
system \eqref{eq:RBF-linear-system} contains $N+1$ unknowns%
\footnote{Namely, we have to determine the $N+1$ parameters $c_{1},\cdots,c_{N}$
and $d$.%
} but only $N$ equations, it is solved using pseudo-invertion \citep[see the Appendix in][]{Zhai2004}.

In contrast to thin-plate splines, the Gaussian function $e^{-\frac{m}{\pi^{2}}S(\mathbf{r},\mathbf{r}_{i})}$
approaches zero asymptotically with growing the distance from the
centers $\mathbf{r}_{i}$. The parameter $m$, commonly referred to
as the \emph{spread parameter}, controls the rate of decay of $e^{-\frac{m}{\pi^{2}}S(\mathbf{r},\mathbf{r}_{i})}$.
Large values of $m$ produce a sharp decay, resulting in a small range
of influence for each node $\mathbf{r}_{i}$. Yao recommended to set
$m$ according to the number of electrodes in the montage; for instance,
$m=20$ for arrays with 32 sensors, $m=40$ for 64 sensors, and $m=50$
for 128 sensors. But he also remarked that a proper choice of $m$
should take into account the source location. In principle, small
values of $m$ would be more suitable to fit deep brain sources, whereas
shallow sources would be better described by small $m$-values. In
other words, interpolation with Gaussian functions is not automatic.
The parameter $m$ needs to be tuned properly for good performance,
which can be difficult to achieve for non-equidistant electrodes.
Nevertheless, Yao showed consistent results favoring the Gaussian
method against spherical splines for simulated and empirical data.
More recently \citet{Bortel2013} has questioned the regularization
technique used by \citet{yao_high-resolution_2002} and the impossibility
of obtaining good estimates without a prior knowledge about the source
depth.

The method developed by Nunez and collaborators is called the \emph{New
Orleans Spline-Laplacian} \citep{law_high-resolution_1993,nunez_electric_2006}.
This method uses an interpolating function that resembles two-dimensional
splines but with knots in $\mathbb{R}^{3}$ instead of $\mathbb{R}^{2}$,
i.e., 
\begin{eqnarray}
f_{\mathrm{NOSL}}(\mathbf{r}) & = & \sum_{i=1}^{N}c_{i}\left|\mathbf{r}-\mathbf{r}_{i}\right|{}^{4}\log(\left|\mathbf{r}-\mathbf{r}_{i}\right|{}^{2}+\omega^{2})+d_{1}+d_{2}x+d_{3}y\nonumber \\
 & + & d_{4}z+d_{5}x^{2}+d_{6}xy+d_{7}xz+d_{8}y^{2}+d_{9}yz+d_{10}z^{2},\label{eq:Other_Approaches-NOSL}
\end{eqnarray}
where the parameters $c_{i}$'s and $d_{i}$'s are determined in the
fashion of equation\,\eqref{eq:TPS_Linear_System}. This method was
implemented for spherical and ellipsoidal scalp models and its performance
was studied using simulations and real data \citep{law_high-resolution_1993}.
A remark about the function $f_{\mathrm{NOSL}}(\mathbf{r})$ is that
it has no optimality property, except that for $\omega$ and $z$
equal to zero it corresponds to the unique minimizer of \eqref{eq:TPS_Variational_Problem}
in $\mathbb{R}^{2}$ \citep{duchon_splines_1977,meinguet_1979}. But
$f_{\mathrm{NOSL}}(\mathbf{r})$ does not minimize \eqref{eq:TPS_Variational_Problem}
in $\mathbb{R}^{3}$, as the minimizer of \eqref{eq:TPS_Variational_Problem}
in $\mathbb{R}^{3}$ is unique and correspond to the spline function
\eqref{eq:TPS_Interpolating_Function}. Or, putting it in other words,
$f_{\mathrm{NOSL}}(\mathbf{r})$ is\emph{ }not actually a spline function.%
\footnote{Due to the constraint \eqref{eq:TPS_Ellipsoid}, the NOSL algorithm
is also affected by the singularity of \eqref{eq:TPS_Linear_System}
on spherical and ellipsoidal scalp models (see discussion in Section\,\ref{sub:Thin-Plate-Splines}).
Apparently, in a effort to remedy this problem the Matlab code of
\citet[p. 585]{nunez_electric_2006} adds ``noise'' (error) to zero-valued
electrode coordinates, effectively making \eqref{eq:TPS_Linear_System}
nonsingular, but not necessarily leading to a well-conditioned solution.%
} Despite that, Nunez and colleagues reported many studies showing
a good performance of their method \citep{law_improving_1993,nunez1994,nunez_2001,nunez_electric_2006}
and emphasized its agreement with cortical imaging algorithms in terms
of correlation \citep{nunez_comparison_1993,nunez_theoretical_1994}.

\subsection{The surface Laplacian matrix \label{sec:The-L-Matrix}}

The computational cost of estimating the surface Laplacian is usually
not prohibitive, but in the majority of circumstances it is high enough
to justify the use of an additional technique to improve performance.
The \emph{surface Laplacian matrix}, denoted by $\mathbf{L}$, is
an excellent tool for this purpose, and because of the linearity of
the Laplacian differentiation, it can be created for any of the Laplacian
methods discussed in this paper. For interpretation purposes the surface
Laplacian matrix can be viewed as a discrete representation of the
differential operator $\mathrm{Lap}_{S}$ at the electrode sites.
We will illustrate the method for finite difference and splines. Once
the matrix $\mathbf{L}$ is obtained the Laplacian estimate is carried
out through the linear transformation 
\begin{equation}
\mathrm{Lap}_{S}(\mathbf{v})=\mathbf{L}\mathbf{v}.\label{eq:Laplacian_transformation}
\end{equation}
An important aspect of $\mathbf{L}$ is that its construction does
not use brain data, as otherwise it could not represent the mathematical
operator $\mathrm{Lap}_{S}$. This way we need to create $\mathbf{L}$
only once for each method and electrode configuration. But notice
that $\mathbf{L}$ depends on the parameters $m$ and $\lambda$ for
splines.

We will first illustrate the construction of $\mathbf{L}$ for finite
difference approximations. In this case, $\mathbf{L}$ is constructed
row by row using the scheme of Figure\,\ref{fig:Five-point-stencils}.
Electrodes that do not contribute to an approximation are assigned
a zero weight in that computation. This way, each row of $\mathbf{L}$
has as many elements as the number of channels in the montage, but
only five elements are nonzero in each row. This implies that $\mathbf{L}$
is a sparse matrix, with its sparsity increasing with the number of
channels. Because the weights of each approximation sum to zero, the
columns of $\mathbf{L}$ also sum to zero, which reflects the fact
that the Laplacian transformation is reference-free.

\begin{figure}
\noindent \centering{}\includegraphics[width=0.3\linewidth]{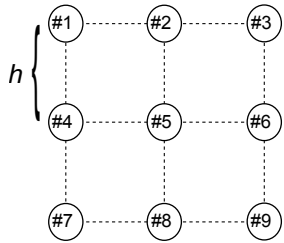}\protect\caption{A montage with 9 electrodes in a square grid of spacing $h$. The
numbers 1 to 9 identify the electrodes and should not be confused
with the weights for a finite difference approximation. \label{fig:montage-with-9-electrodes}}
\end{figure}

As a simple illustration, consider the electrode arrangement of Figure\,\ref{fig:montage-with-9-electrodes},
with 9 electrodes evenly distributed in a square grid of spacing $h$.
The Laplacian matrix of this arrangement has dimension $9\times9$
and is given by 
\begin{equation}
\mathbf{L}=\frac{1}{h^{2}}\begin{pmatrix}\begin{array}{rrrrrrrrr}
2 & -2 & 1 & -2 & 0 & 0 & 1 & 0 & 0\\
1 & -1 & 1 & 0 & -2 & 0 & 0 & 1 & 0\\
1 & -2 & 2 & 0 & 0 & -2 & 0 & 0 & 1\\
1 & 0 & 0 & -1 & -2 & 1 & 1 & 0 & 0\\
0 & 1 & 0 & 1 & -4 & 1 & 0 & 1 & 0\\
0 & 0 & 1 & 1 & -2 & -1 & 0 & 0 & 1\\
1 & 0 & 0 & -2 & 0 & 0 & 2 & -2 & 1\\
0 & 1 & 0 & 0 & -2 & 0 & 1 & -1 & 1\\
0 & 0 & 1 & 0 & 0 & -2 & 1 & -2 & 2
\end{array}\end{pmatrix}.\label{eq:SL_Matrix-Example-of-L}
\end{equation}
A quick inspection on the diagonal elements of $\mathbf{L}$ reveals
that none of them is equal to zero, but they sum to zero. This means
that $\mathbf{L}$ \emph{does not possess inverse}. The inability
to invert $\mathbf{L}$ is not exclusive of finite differences, but
rather a general property that reflects the ambiguity on the choice
of reference. Stated in other words, \emph{the Laplacian transformation
\eqref{eq:Laplacian_transformation} cannot be uniquely undone due
to the fact that the potential is not uniquely defined}.

Mesh-free methods generate $\mathbf{L}$ matrices that are non-sparse,
i.e., most of their entries are nonzero. In the context of smoothing
splines, \citet{Carvalhaes2011} showed that 
\begin{equation}
\mathbf{L}_{\lambda}=\tilde{\mathbf{K}}\,\mathbf{C}_{\lambda}+\tilde{\mathbf{T}}\,\mathbf{D}_{\lambda},\label{eq:SL_Matrix-differentiation-matrix}
\end{equation}
where the subscript $\lambda$ emphasizes dependency on the regularization
parameter, and\begin{subequations}\label{eq:SL_Matrix-Matrices_C_and_D}
\begin{gather}
(\tilde{\mathbf{K}})_{ij}=\mathrm{Lap}_{s}\left(\left|\mathbf{r}_{i}-\mathbf{r}_{j}\right|^{2m-3}\right),\\
(\tilde{\mathbf{T}})_{ij}=\mathrm{Lap}_{s}\left(\nu_{i}(\mathbf{r}_{j})\right),\\
\mathbf{C}_{\lambda}=\mathbf{Q}_{2}\left[\mathbf{Q}_{2}^{T}\left(\mathbf{K}+N\lambda\mathbf{I}\right)\mathbf{Q}_{2}\right]^{-1}\mathbf{Q}_{2}^{T},\label{eq:C_lambda}\\
\mathbf{D}_{\lambda}=\mathbf{R}^{+}\mathbf{Q}_{1}^{T}\left(\mathbf{1}-\mathbf{K}\mathbf{C}_{\lambda}-N\lambda\mathbf{C}_{\lambda}\right).
\end{gather}
\end{subequations}The matrix $\mathbf{L}_{\lambda}$ also has no
inverse, thus not allowing unambiguous inverse operation. This is
due to the fact that $\mathbf{C}_{\lambda}$ and $\tilde{\mathbf{T}}$
are rank-deficient. In other words, since $\mathbf{Q}_{2}$ has only
$N-1$ columns, only $N-1$ degrees of freedom are available from
$\mathbf{C}_{\lambda}$. In turn, because the Laplacian operation
$\mathrm{Lap}_{s}(\nu_{i}(\mathbf{r}_{j}))$ vanishes at any polynomial
$\nu_{i}$ of degree less than 2, the first four columns of $\tilde{\mathbf{T}}$
are null.

For spherical splines, $(\mathbf{\tilde{\mathbf{K}}})_{ij}=-g_{m}(\mathbf{r}_{j},\mathbf{r}_{i})/r^{2}$,
$\mathbf{T}$ is a $N$-vector of ones, and $\tilde{\mathbf{T}}$
is null, so that 
\begin{equation}
\mathbf{L}_{\lambda}=\tilde{\mathbf{K}}\,\mathbf{C}_{\lambda},
\end{equation}
which has rank $N-1$ and does not possess inverse. Since $\mathbf{T}$
is a vector of ones, its QR-factorization gives $\mathbf{R}=-\sqrt{N}$,
i.e., $\mathbf{R}$ is a scalar instead of a matrix, and $\mathbf{Q}_{1}=-\mathbf{T}/\sqrt{N}$.
Consequently, $\mathbf{D}_{\lambda}=\mathbf{T}^{T}\left(\mathbf{1}-\mathbf{K}\mathbf{C}_{\lambda}-N\lambda\mathbf{C}_{\lambda}\right)/N$,
which is also a $N$-vector instead of a matrix.

As anticipated, the construction of $\mathbf{L}$ and $\mathbf{L}_{\lambda}$
is carried out without using brain data. Because the matrix $\mathbf{C}_{\lambda}$
depends on the inverse of the matrix $[\mathbf{Q}_{2}^{T}\left(\mathbf{K}+N\lambda\mathbf{I}\right)\mathbf{Q}_{2}]$
it is subject to the numerical issues discussed in Section~\ref{sub:Thin-Plate-Splines},
so that it is a good measure to set $\lambda$ different from zero
and keep $m$ as small as possible for accurate results.

The computational cost of estimating the Laplacian can be expressed
in terms of the number of floating-point operations (flops) required
by \eqref{eq:Laplacian_transformation}. The computational cost of
finite difference is generally much smaller than the cost of splines.
Assume that $\mathbf{v}$ in \eqref{eq:Laplacian_transformation}
is a matrix containing $N$ rows, each one corresponding to an electrodes,
and $T$ columns, which one representing a time frame. The cost of
computing \eqref{eq:Laplacian_transformation} using finite difference
is $5NT$~flops%
\footnote{This formula is valid only if $\mathbf{L}$ is stored in the computer
as a sparse matrix.%
}, which is equivalent to the number of nonzero elements in $\mathbf{L}$
times the number of time frames. Putting in numbers, handling 5~minutes
of recordings at 1\,kHz sampling rate costs $5\times32\times18000=2,880,000$~flops
for $N=32$~electrodes. This cost is much higher in the spline framework.
Because $\mathbf{L}_{\lambda}$ is not sparse, the cost of computing
\eqref{eq:Laplacian_transformation} is $N^{2}T$~flops, or 18,432,000~flops
for the above example. However, without using the Laplacian matrix
the system \eqref{eq:TPS_Linear_System} needs to be solved once for
each frame at the cost of $(2/3)~N^{3}$~flops%
\footnote{This formula is valid for a solution using LU decomposition \citep{press_numerical_1992}.%
} per frame, which brings the figure to over 6~billion flops.

\subsection{The regularization of smoothing splines\label{sec:Data-Smoothing}}

The use of smoothing splines to estimate the surface Laplacian requires
the estimate of the regularization parameter $\lambda$, which, as
said above, has the beneficial effect of eliminating spatial noise
through variance reduction. A key factor to guide towards a good choice
of $\lambda$ is the mean square error function 
\begin{equation}
R(\lambda)=\frac{1}{N}\sum_{i=1}^{N}\left(V_{i}-f_{\lambda}(\mathbf{r}_{i})\right)^{2}.\label{eq:square_error}
\end{equation}
Intuitively, this function measures infidelity of $f_{\lambda}(\mathbf{r})$
to the input data ($V_{1},\cdots,V_{N}$) due to smoothing. The goal
of regularization is to weight this function appropriately to achieve
a good balance between fidelity and variance reduction. The popular
method of \emph{generalized cross-validation} (GCV) proposed by \citet{Craven1979}
addresses this problem in the following way. The GCV estimate of $\lambda$
is the value for which the function 
\begin{equation}
\mathrm{GCV}(\lambda)=\frac{1}{N}\frac{\sum_{i=1}^{N}\left(V_{i}-f_{\lambda}(\mathbf{r}_{i})\right)^{2}}{\left(1-\mathrm{trace}(\mathbf{S}_{\lambda})/N\right)^{2}},\label{eq:GCV_function}
\end{equation}
reaches its minimum value. Here, $\mathbf{S}_{\lambda}$ is a $N\times N$
matrix satisfying
\begin{equation}
\begin{pmatrix}f_{\lambda}(\mathbf{r}_{1})\\
\vdots\\
f_{\lambda}(\mathbf{r}_{N})
\end{pmatrix}=\mathbf{S}_{\lambda}\begin{pmatrix}V_{1}\\
\vdots\\
V_{N}
\end{pmatrix}.\label{eq:Smoothing-The_smooth_matrix}
\end{equation}
In other words, the GCV criterion corrects the squared residuals $\left(V_{i}-f_{\lambda}(\mathbf{r}_{i})\right)^{2}$
about the estimate by dividing them by the factor $(1-\mathrm{trace}(\mathbf{S}_{\lambda})/N)^{2}$.
The value $\lambda=0$ is excluded from predictions because for $\lambda=0$
\eqref{eq:Smoothing-The_smooth_matrix} is turned into a regression
problem for which $\mathrm{trace}(\mathbf{S}_{\lambda})=N$, which
vanishes the denominator of \eqref{eq:GCV_function}. 

The matrix $\mathbf{S}_{\lambda}$ is called the \emph{smoother matrix.}
It follows from our previous development that\emph{ }
\begin{equation}
\mathbf{S}_{\lambda}=\mathbf{K}\,\mathbf{C}_{\lambda}+\mathbf{T}\,\mathbf{D}_{\lambda}.\label{eq:the_smoother_matrix}
\end{equation}
Since this expression involves no differentiation, it is relatively
simple to compute $\mathbf{S}_{\lambda}$ on any scalp model. Another
observation is that the columns of $\mathbf{S}_{\lambda}$ sum to
1 (whereas the columns of $\mathbf{L}_{\lambda}$ sum to 0), implying
that the transformation \eqref{eq:Smoothing-The_smooth_matrix} preserves
the reference potential.

In practice, the search for $\lambda$ is constrained to a finite
interval which typically spans several decades on a log scale \citep{babiloni_performances_1995}.
In order to facilitate interpretation and comparisons across studies,
instead of thinking in terms of $\lambda$ it is convenient to think
in terms of a parameter $\mathrm{DF}_{\lambda}$ that has more practical
meaning. The parameter $\mathrm{DF}_{\lambda}$ is called the\emph{
effective degree of freedom} of $f_{\lambda}(\mathbf{r})$ and is
defined by \citep{Hastie2009,James2013} 
\begin{equation}
\mathrm{DF}_{\lambda}=\mathrm{trace}(\mathbf{S}_{\lambda}),\label{eq:degrees_of_freedom}
\end{equation}
where the trace operation sums over the diagonal elements of $\mathbf{S}_{\lambda}$.
This operation assigns a real value to $\mathrm{DF}_{\lambda}$ between
1 and $N$.%
\footnote{The lower bound $\mathrm{DF}_{\lambda}=1$ holds for spherical splines.
For regular splines $\mathrm{DF}_{\lambda}$ is at least equal to
2, depending on the dimension of the space. %
} The lower bound of $\mathrm{DF}_{\lambda}$ accounts for the linear
portion of $f_{\lambda}(\mathbf{r})$, whereas the upper bound corresponds
to $\lambda=0$, for which $\mathbf{S}_{\lambda}$ is the projection
matrix from regression over $(V_{1},\cdots,V_{N})^{T}$ which has
trace equal to $N$.

\begin{figure}
\noindent \centering{}\includegraphics[width=1\textwidth]{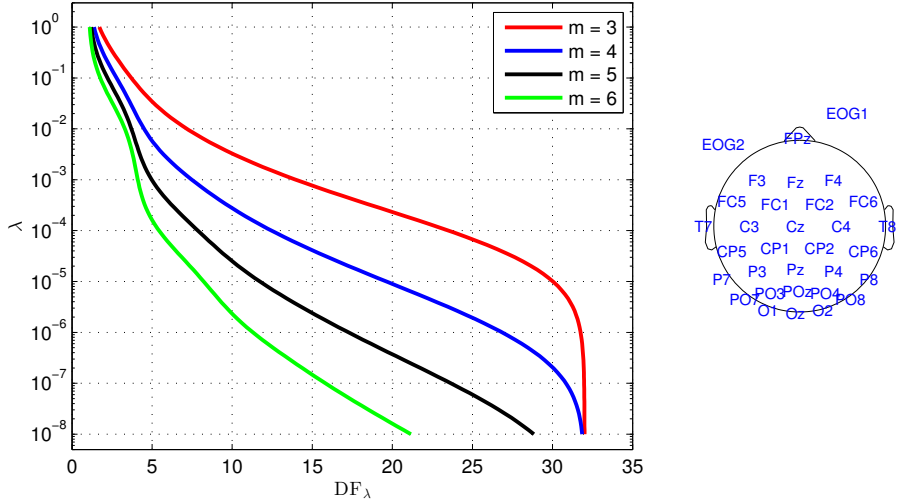}\protect\caption{The monotonically decreasing relationship between
$\mathrm{DF}_{\lambda}$ and $\lambda$ for a montage with 32 electrodes
(right). Each solid line indicates a distribution with 100 values
of $\lambda$ in the range of $10^{-8}$ to 1 on a log scale. The
different colors indicate the value of the parameter $m$ used in
the calculation.\label{fig:GCV-DF-Lambda}}
\end{figure}

In order to think in terms of $\mathrm{DF}_{\lambda}$ instead of
$\lambda$ we need to invert the relation \eqref{eq:degrees_of_freedom}.
This inversion is possible but it can only be done numerically. Figure~\ref{fig:GCV-DF-Lambda}
depicts the situation for a montage with 32 sensors. Although $\mathrm{DF}_{\lambda}$
is shown as if it were the independent variable, its values were actually
determined from \eqref{eq:degrees_of_freedom} by fixing $\lambda$.
The Matlab code of Appendix~B was used to compute $\mathbf{S}_{\lambda}$.
This code implements the spherical splines algorithm of Section~\ref{sub:Spherical-Splines}
with a default error tolerance of less than $1\times10^{-10}$ in
the computation of the infinity series \eqref{eq:SS_Kernel}. The
electrode locations were normalized to a sphere of radius 10~cm.
Figure~\ref{fig:GCV-DF-Lambda} shows that $\mathrm{DF}_{\lambda}$
decreases monotonically with $\lambda$. Thanks to monotonicity we
can interpolate the pairs $(\mathrm{DF}_{\lambda},\lambda)$ to obtain
a continuous curve from which we can estimate $\lambda$ given $\mathrm{DF}_{\lambda}$.

To fix ideas lets consider an example with real brain data. The data
that will be used here were extracted from a sample dataset distributed
with the EEGLab Matlab Toolbox, version 13.2.2b \citep{Delorme2004}.
Our goal is simply to illustrate the regularization procedure in a
way that can be easily reproduced by the reader. The ready availability
of this dataset is very suitable for this purpose. The data contain
30,504 frames recorded from 32 channels at a sampling rate of 128\,Hz.
We arbitrarily chose the frame of number 200 for exploration. Our
computations were carried out using spherical splines with $m=4$
(see Appendix~B). The electrode configuration was the same as in
Figure~\ref{fig:GCV-DF-Lambda}. The following steps were followed:
\begin{enumerate}
\item Create a grid of values of $\mathrm{DF}_{\lambda}$ between 2 and
31 in increments of 1.
\item Calculate the values of $\lambda$ from $\mathrm{DF}_{\lambda}$ by
means of interpolation.
\item Compute the GCV score of each pair $\mathrm{DF}_{\lambda}$ and $\lambda$
from \eqref{eq:GCV_function}.
\item Interpolate the pairs $(\mathrm{DF}_{\lambda},\mbox{GCV})$ to obtain
a continuous error curve.
\item Find the optimal value $\hat{\mathrm{DF}}_{\lambda}$ that minimizes
the error curve.
\end{enumerate}
Step 1 is the only step of this sequence that depends on the electrode
configuration, with the upper bound \LyXZeroWidthSpace \LyXZeroWidthSpace of
$\mathrm{DF}_{\lambda}$ depending on the number of electrodes. The
Matlab function \texttt{interp1} was used to perform step~2, whereas
steps~4 and 5 were carried out using \texttt{csapis, fnval, and fminbnd}. 

\begin{figure}
\noindent \centering{}\includegraphics[width=1\textwidth]{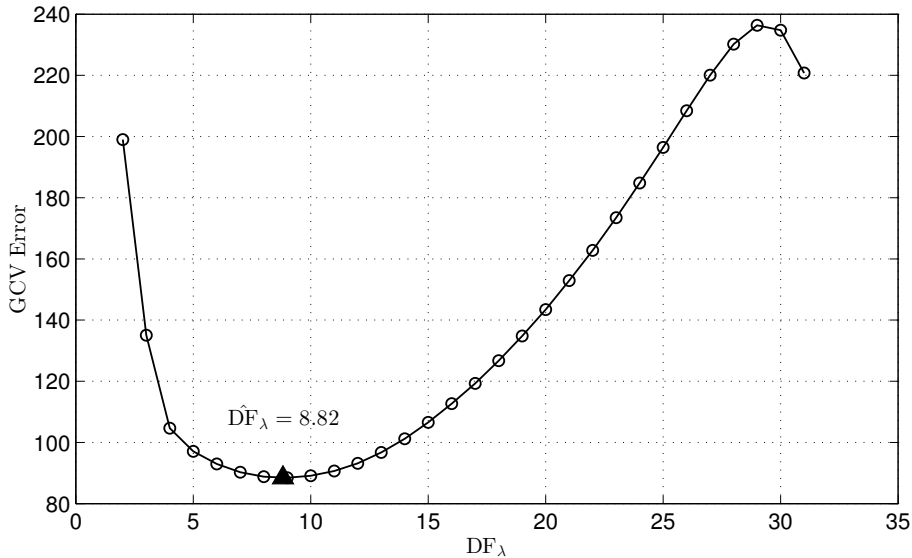}\protect\caption{An illustration of the GCV regularization. The error
curve reached its minimum value at $\hat{\mathrm{DF}}_{\lambda}=8.82$.}\label{fig:GCV}
\end{figure}

\begin{figure}
\centering{}\includegraphics[width=1\textwidth]{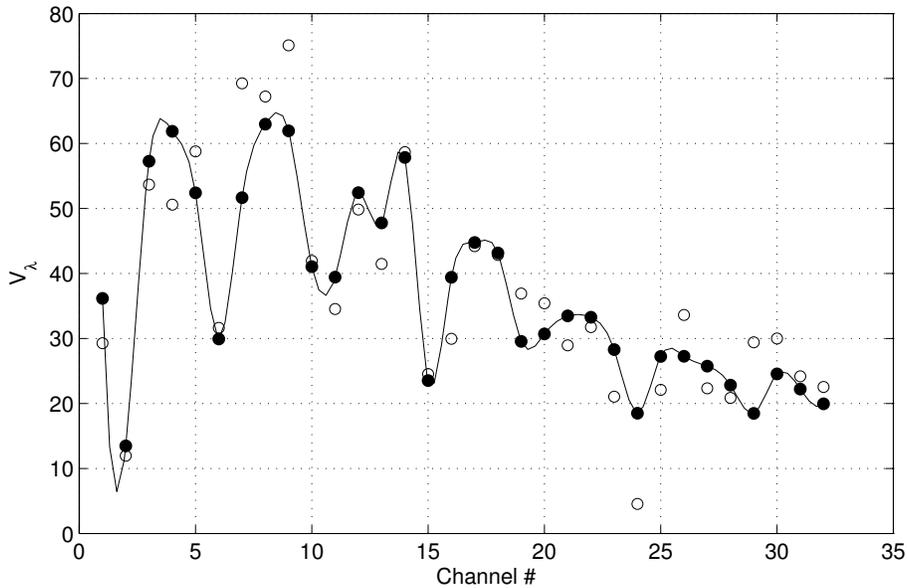}\protect\caption{Spline fit (filled circles) of the sample potential
(empty circles) using $\mathrm{DF}_{\lambda}=8.82$. The solid line
is only a guide to the eye.} \label{fig:GCV-the-smooth-waveform}
\end{figure}

The error curve is depicted in Figure\,\ref{fig:GCV}. The optimal
value of $\mathrm{DF}_{\lambda}$ was $\hat{\mathrm{DF}}_{\lambda}=8.82$,
corresponding to a reduction by about 72\% in degrees of freedom.
Figure\,\ref{fig:GCV-the-smooth-waveform} shows a plot of the smooth
potential for this value of $\hat{\mathrm{DF}}_{\lambda}$. Note an
important change in topography due to regularization. While the raw
potential exhibits two prominent peaks at electrodes 7 (FC5) and 9
(FC2), which suggests the presence of a pair of dipoles of same polarity
interposed by FC1, the regularized potential points out the existence
of a single dipole in the region enclosed by Fz, FC1, FC2, and Cz
(electrodes 4, 8, 9, and 14). 

In this example the potential distribution was re-referenced to the
average before regularization, but because the smoother matrix preserves
the reference, both before and after regularization it had zero mean,
as can be inferred from Figure\,\ref{fig:GCV-the-smooth-waveform}.
We also remark that the theory underlying GCV is asymptotic, so that
a few outliers should be expected for $N$ small. For this reason
the error function should be minimized by constraining the solution
to the extremes of the interval in step~1.

We will now focus on a problem of more practical interest, which is
the simultaneous regularization of multiple frames. In principle,
the above procedure applies immediately if we construct the spline
function in $\mathbb{R}^{4}$ instead of $\mathbb{R}^{3}$ by taking
time as a fourth coordinate. However, it is not difficult to see that
this cannot work in practice, because of dimensionality. For instance,
a data sample with 32 channels as above and just 100 frames sets the
dimension of the system \eqref{eq:TPS_Linear_System} to 3,200$\times$3,200,
which is far beyond the limit of validity of splines. Due to this
problem EEG systems containing more than 100 channels are unfeasible
to be dealt with splines in $\mathbb{R}^{4}$ even for a small number
of frames. For $N$ small we could think of slicing the data into
small non-overlapping sets for regularization, but it is questionable
if this can improve to much on frame-by-frame regularization in $\mathbb{R}^{3}$.
So, let us continue to work with splines in $\mathbb{R}^{3}$.

Although the GCV algorithm is relatively simple and requires modest
computational resources and time, its sequential application over
multiple frames can be costly. In addition, it is very unlikely that
outcomes from independent regularizations would agree upon a global
estimate of $\hat{\mathrm{DF}}_{\lambda}$, and dealing with a distribution
of $\hat{\mathrm{DF}}_{\lambda}$ lacks flexibility and simplicity
sought so far. Of course, we could average across GCV estimates to
determine a global $\hat{\mathrm{DF}}_{\lambda}$, but there is no
reason to believe that this distribution is stationary and has a central
tendency. Viewed from a statistical perspective, it would be more
interesting to seek an optimality criterion that applies at once to
the entire dataset. In light of this commitment we suggest a modification
to the GCV algorithm, expressed as follows 
\begin{equation}
\overline{\mathrm{GCV}}(\lambda)=\frac{\sum_{i=1}^{N}\sum_{t=1}^{T}\left(V_{i,t}-f_{\lambda,t}(\mathbf{r}_{i})\right)^{2}}{NT\left(1-\mathrm{DF}_{\lambda}/N\right)^{2}},\label{eq:GCV_function2}
\end{equation}
where $T$ is the total number of frames and $V_{i,t}$ and $f_{\lambda,t}$
are the potential and the estimated spline function on frame $t$.
That is, each value of $\mathrm{DF}_{\lambda}$ is now scored according
to its global effect across all frames rather than on a single frame.
The weight $1/\left(1-\mathrm{DF}_{\lambda}/N\right)^{2}$ assigned
to each choice of $\mathrm{DF}_{\lambda}$ remains the same, but now
closeness of fit as measured by mean square error is influenced by
all residuals at once, which justifies the term $1/NT$ to normalize
the square error. Another way of putting it is that the $\overline{\mathrm{GCV}}$
estimate of $\hat{\mathrm{DF}}_{\lambda}$ is the value of $\mathrm{DF}_{\lambda}$
that minimizes the mean GCV error function $\overline{\mathrm{GCV}}(\lambda)=\sum_{t=1}^{T}\mathrm{GCV}_{t}(\lambda)/T$.

\begin{figure}
\noindent \begin{centering}
\includegraphics[width=1\textwidth]{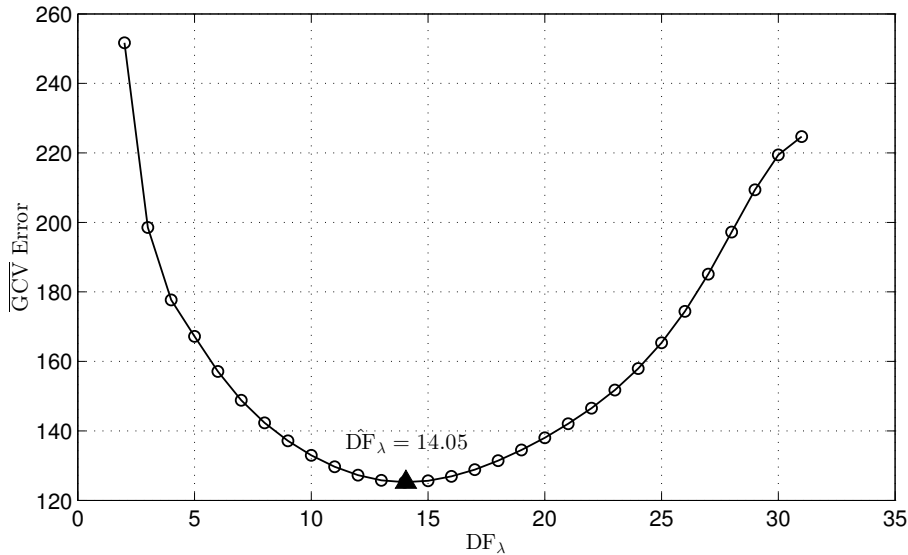}
\par\end{centering}

\protect\caption{Global regularization using the $\overline{\mathrm{GCV}}$
criterion.}\label{fig:GCV2}
\end{figure}

We applied this criterion to regularize the above dataset for all
frames. Except that equation~\eqref{eq:GCV_function2} instead of
\eqref{eq:GCV_function} was used to compute error scores, the same
5-step procedure of the previous example was followed. Figure \ref{fig:GCV2}
depicts the error curve. The estimated value of $\hat{\mathrm{DF}}_{\lambda}$
is $14.05$, which differs significantly from the previous example.

\section{Concluding Remarks}

The literature on the surface Laplacian technique is quite large and
this paper could only partially review or expand a few topics of general
importance. It was our primary goal to give a more intuitive view
of the technique by providing physical insights that are often missing
in the literature. In addition, we discussed several numerical methods
to estimate the Laplacian, each one has its own set of strengths and
limitations which we tried to clarify. Because of popularity, special
attention was given to the methods of finite difference and splines.
Finite difference is arguably one of the simplest approach for Laplacian
estimates and its computational cost is highly encouraging, mainly
for EEG experiments involving a large number of electrodes. The major
disadvantage, however, is that estimates are prone to discretization
errors and regularized estimates to eliminate spatial noise are not
provided. Splines are still the main alternative to finite difference
in EEG literature. The need for evenly-spaced electrodes and other
problems related to discretization are totally eliminated, the scalp
geometry becomes more realistic, and spatial noise that otherwise
is amplified by Laplacian differentiation are reduced through regularization.
The cost of such improvement is not negligible and is reflected in
a increase in mathematical complexity and a higher computational cost
compared to finite difference. Moreover, it worth remarking that splines
are also sensitive to error in electrode locations \citep[e.g.][]{Bortel2008}
and the belief that accurate estimates at peripheral electrodes are
made possible is not entirely correct. In fact, note that the construction
of $f_{\lambda}(\mathbf{r})$ involves $N+M$ unknowns, which are
the coefficients $c_{i}$'s and $d_{i}$'s in \eqref{eq:TPS_Interpolating_Function}
and \eqref{eq:SS_Interpolatiing_Function}, but only $N$ observations
are available for each frame, which are the measured potentials $V_{1},\cdots,V_{N}$.
To avoid underdetermination, the system \eqref{eq:TPS_Linear_System}
contains an ad hoc condition $\mathbf{T}^{T}\mathbf{c}=0$ that adds
$M$ more constraint to $f_{\lambda}(\mathbf{r})$, which has the
geometric effect of canceling terms from $f_{\lambda}(\mathbf{r})$
that grow faster than a polynomial of degree $m$ as one moves away
from the electrodes. Estimates at central and peripheral electrodes
are affected differently by this condition, with estimates at peripheral
electrodes being less accurate. Unfortunately, it was outside the
scope of our work to discuss methods to make the scalp model more
realistic and reduce error in electrode locations, as they require
additional tools such as MRI that are often not readily available
to most researchers. The interested reader is referred to the relevant
references for this matter \citep{Le1993,gevins_high_1994,le_local_1994,babilon_spline_Laplacian_1996,Bortel2013}.

Our discussion on computational methods emphasized our recommendation
to set the spline parameter $m$ as equal to 3 or 4 to avoid numerical
issues. We can use the result in Figure~\ref{fig:GCV-DF-Lambda}
to get further insight into this problem. This example clearly shows
a compromise between the value of $m$, the effective number of degrees
of freedom of the spline fit, and the lower bound of the $\lambda$
parameter. Namely, although small and large values of $m$ appear
equally suitable to assess low degrees of freedom, say $\mathrm{DF}_{\lambda}<10$,
the fact that $\lambda$ is bounded from below at $10^{-8}$ prevents
us from exploiting the region of $\mathrm{DF}_{\lambda}>20$ properly
without setting $m=3$ or 4. One may think of a way to circumvent
this limitation by decreasing the lower bound of $\lambda$ toward
zero, but as explained in Section~\ref{sub:Thin-Plate-Splines} this
has the problem of impoverishing the numerical conditioning, thus
reinforcing our recommendation stated above.

Yet, despite the broad literature on the subject, some basic problems
on Laplacian estimates are still unsolved. Our effort to contribute
to such questions is reflected in a new set of approximations for
Laplacian estimates on the grid and a workable solution to the problem
of global regularization. One difficult problem that could not be
addressed is related to the finite size of measuring electrodes. All
computational methods discussed here were built upon the idea of a
point-like electrode, but an actual electrode has a finite size and
for this reason EEG potentials should be interpreted as spatial averages
instead of point values. From a computational perspective, a similar
problem occurs in the approximation of a histogram by a continuous
function in statistics (the density function), which has led to the
concept of \emph{histopolation} \citep{Boneva1971,Wahba1976}. The
solution consisted basically of replacing the point interpolation
condition $y_{i}=f(x_{i})$ by the \emph{area matching condition}
$y_{i}=\int_{ih}^{(i+1)h}f(x)\thinspace dx$, where $h$ is the ``bin
size''. An extension to two-dimensional space was straightforward
and has led to the variational problem of minimizing 
\begin{equation}
\mbox{RSS}(f,\lambda)=\frac{1}{N}\sum_{i}\left[V_{i}-\iint_{\Omega_{i}}f(\mathbf{r}_{i})\thinspace dA\right]^{2}+\lambda\thinspace J_{m}\left[f\right].\label{eq:histopolation}
\end{equation}
\citet{Wahba1981} studied this problem in Euclidean space and used
the GCV statistics as explained here to estimate the optimal parameter
$\lambda$. A similar approach could be sought on the sphere to account
for finite-sized electrodes on a spherical scalp model, taking the
double integral over the finite area of each electrode. This problem
is arguably not trivial and almost certainly will require a numerical
solution. Yet this a simplified model that does not address the much
more complicate problem of boundary effects due to the highly conductive
electrode over the region $\Omega_{i}$.

Considerable attention was devoted to the method of estimating the
Laplacian derivation by means of a linear transformation, for which
some figures were given estimating computational costs and encouraging
the use of this procedure, mainly in the context of splines. Similar
attention was paid to the problem of regularization across multiple
frames to improve estimates, for which we suggested a possible solution
based on generalized cross-validation. It is worth remarking that
spatial regularization is not useful to eliminate noise coming from
the reference electrode. In fact, because spline fits preserves the
reference signal, the residuals $\varepsilon_{i}=V_{i}-f_{\lambda}(\mathbf{r}_{i})$
are reference-free, and so are the error functions GCV and $\overline{\mathrm{GCV}}$,
making evident that any noise due to the reference is fully preserved
by regularization. This problem, however, should not cause any concern
since it simply does not exist in the context of surface Laplacian
analysis.

Finally, we want to mention our effort in recent papers to develop
a vector form of EEG that combines the surface Laplacian and the tangential
components of the electric field projected on the scalp surface \citep{Wang2012,Carvalhaes2014}.
This combination was grounded on the observation that the surface
Laplacian derivation is closely related to the spatial component of
the electric field oriented normally to the scalp surface, thus not
containing substantial information about features encoded in tangential
direction. Not surprisingly, this combination of physically distinct
but somewhat supplementary quantities resulted in significant improvement
for a variety of classification tasks as described in \citet{Carvalhaes2014}.

\appendix

\section{The mathematics Behind Finite Difference}

A systematic way to obtain finite difference approximations for the
derivatives of a function is based on the \emph{Taylor series expansion}.
To expand in terms of Taylor series, we formulate the problem in terms
of a continuous function $V$ and then, for compatibility with our
notation in Section\,\ref{sub:Finite-Differences}, rewrite the final
approximation using discrete variables. In its simplest case, the
Taylor series of a univariate function $V(x)$ around a point $x=a$
is defined by the infinite summation 
\begin{equation}
V(a+h)=V(a)+V^{\prime}(a)h+\frac{1}{2!}V^{\prime\prime}(a)h^{2}+\frac{1}{3!}V^{\prime\prime\prime}(a)h^{3}+\cdots,\label{eq:one_dimensional_Taylor_series}
\end{equation}
where $h$ is called the increment, $n!=n(n-1)(n-2)\cdots1$ is the
factorial of $n$, and $V^{(n)}\left(a\right)$ denotes the $n$-th
derivative of $V(x)$ at $x=a$. That is, the Taylor series of $V(x)$
is a power series of $h$ with coefficients given by the derivatives
of $V(x)$ at $x=a$. To obtain an approximation for the second derivative
(or any other derivative) of $V(x)$ at $x=a$ we proceed as follows.
Replace $h$ with $-h$ in \eqref{eq:one_dimensional_Taylor_series}
and obtain an analogous expression for $V(a-h)$, i.e., 
\[
V(a-h)=V(a)-V^{\prime}(a)h+\frac{1}{2!}V^{\prime\prime}(a)h^{2}-\frac{1}{3!}V^{\prime\prime\prime}(a)h^{3}+\cdots
\]
Adding these two expressions, 
\[
V(a+h)+V(a-h)=2V(a)+V^{\prime\prime}(a)h^{2}+\frac{1}{12}V^{iv}(a)h^{4}+\cdots,
\]
whence 
\begin{equation}
V^{\prime\prime}(a)=\frac{V(a+h)+V(a-h)-2V(a)}{h^{2}}-\frac{1}{12}V^{iv}(a)h^{2}-\cdots,\label{eq:appendix-second_derivative_of_f}
\end{equation}
Assuming that $h$ is sufficiently small we can ignore all terms containing
powers of $h$ and arrive at 
\[
V^{\prime\prime}(a)\approx\frac{V(a+h)+V(a-h)-2V(a)}{h^{2}}.
\]
Replacing $a=(i-1)\, h$, $i=2,\cdots,N-1$, and using the notation
$V_{i}=V(a)$ and $V_{i\pm1}=V(a\pm h)$ yields 
\begin{equation}
V^{\prime\prime}(a)\approx\frac{V_{i+1}+V_{i-1}-2V_{i}}{h^{2}}.\label{eq:appendix-1D-central_node}
\end{equation}
This expression was obtained by truncating the infinite series \eqref{eq:appendix-second_derivative_of_f}
after the first term. This way, starting from $h^{2}$ all powers
of $h$ were eliminated assuming that $h$ was sufficiently small.
The result was a \emph{second-order approximation}%
\footnote{The order of a finite difference approximation is given by the smallest
power of $h$ that is left out in the truncation.%
}, which in the ``big-Oh'' notation is referred to as ${\cal O}(h^{2})$.

Approximation \eqref{eq:appendix-1D-central_node} excludes the case
of a peripheral node, for which either $V_{i+1}$ or $V_{i-1}$ does
not exist. In this case, we have to build the approximation using
only nodes to the right or left of $a$, but not both. For instance,
let $i=1$ and use the expansions 
\begin{gather*}
V(a+h)=V(a)+V^{\prime}(a)h+\frac{1}{2}V^{\prime\prime}(a)h^{2}+\frac{1}{6}V^{\prime\prime\prime}(a)h^{3}+\cdots,\\
V(a+2h)=V(a)+2V^{\prime}(a)h+2V^{\prime\prime}(a)h^{2}+\frac{4}{3}V^{\prime\prime\prime}(a)h^{3}+\cdots,
\end{gather*}
from which 
\[
V(a+2h)-2V(a+h)=-V(a)+V^{\prime\prime}(a)h^{2}+V^{\prime\prime\prime}(a)h^{3}+\cdots
\]
and 
\[
V^{\prime\prime}(a)=\frac{V(a+2h)-2V(a+h)+V(a)}{h^{2}}-V^{\prime\prime\prime}(a)h-\cdots
\]
In terms of the grid variables, 
\begin{equation}
V^{\prime\prime}(a)\approx\frac{V_{3}-2V_{2}+V_{1}}{h^{2}},
\end{equation}
which is a \emph{first order approximation}. Note that this expression
corresponds to the formula obtained by \citet{Freeman1975} for one-dimensional
estimate of intracranial CSD.

\subsection{Bivariate functions and approximations for $\mathrm{Lap}(V(a,b))$}

The extension of \eqref{eq:one_dimensional_Taylor_series} to bivariate
functions replaces ordinary derivatives with partial derivatives.
Namely, 
\begin{multline}
V(a+h_{1},b+h_{2})=V(a,b)+\left[\frac{\partial V(a,b)}{\partial x}h_{1}+\frac{\partial V(a,b)}{\partial y}h_{2}\right]+\frac{1}{2}\left[\frac{\partial^{2}V(a,b)}{\partial x^{2}}h_{1}^{2}+2\frac{\partial^{2}V(a,b)}{\partial x\partial y}h_{1}h_{2}\right.\\
\left.+\frac{\partial^{2}V(a,b)}{\partial y^{2}}h_{2}^{2}\right]+\frac{1}{6}\left[\frac{\partial^{3}V(a,b)}{\partial x^{3}}h_{1}^{3}+3\frac{\partial^{3}V(a,b)}{\partial^{2}x\partial y}h_{1}^{2}h_{2}+3\frac{\partial^{3}V(a,b)}{\partial x\partial^{2}y}h_{1}h_{2}^{2}+\frac{\partial^{3}V(a,b)}{\partial y^{3}}h_{2}^{2}\right]+\cdots\,\,.\label{eq:two_dimensional_Taylor_series}
\end{multline}
Similar to the one-dimensional case, with appropriate choices for
$h_{1}$ and $h_{2}$ one can use this expansion to approximate the
partial derivatives of $V(x,y)$ at any point in a rectangular grid.
An approximation for the Laplacian of $V(x,y)$ at a central node
of a square grid ($h_{1}=h_{2}=h$) is obtained as follows. First,
use \eqref{eq:two_dimensional_Taylor_series} to expand $V(a\pm h,b)$
and $V(a,b\pm h)$: \begin{subequations} 
\begin{align*}
V(a\pm h,b) & =V(a,b)\pm\frac{\partial V(a,b)}{\partial x}h+\frac{1}{2}\frac{\partial^{2}V(a,b)}{\partial x^{2}}h^{2}\pm\frac{1}{6}\frac{\partial^{3}V(a,b)}{\partial x^{3}}h^{3}+\frac{1}{24}\frac{\partial^{4}V(a,b)}{\partial x^{4}}h^{4}\pm\cdots,\\
V(a,b\pm h) & =V(a,b)\pm\frac{\partial V(a,b)}{\partial y}h+\frac{1}{2}\frac{\partial^{2}V(a,b)}{\partial y^{2}}h^{2}\pm\frac{1}{6}\frac{\partial^{3}V(a,b)}{\partial y^{3}}h^{3}+\frac{1}{24}\frac{\partial^{4}V(a,b)}{\partial y^{4}}h^{4}+\cdots.
\end{align*}
\end{subequations}Second, add these four expressions to obtain 
\begin{multline*}
V(a+h,b)+V(a-h,b)+V(a,b+h)+V(a,b-h)=4V(a,b)+\underset{\mathrm{=Lap}(V(a,b))}{\underbrace{\left[\frac{\partial^{2}V(a,b)}{\partial x^{2}}+\frac{\partial^{2}V(a,b)}{\partial y^{2}}\right]}}h^{2}\\
+\frac{1}{12}\left[\frac{\partial^{4}V(a,b)}{\partial x^{4}}+\frac{\partial^{4}V(a,b)}{\partial y^{4}}\right]h^{4}+\cdots,
\end{multline*}
whence 
\[
\mathrm{Lap}(V(a,b))=\frac{V(a+h,b)+V(a-h,b)+V(a,b+h)+V(a,b-h)-4V(a,b)}{h^{2}}+{\cal O}(h^{2}).
\]
Therefore, $\mathrm{Lap}(V(a,b))$ is approximated to the second-order
of $h$ by 
\[
\mathrm{Lap}(V(a,b))\approx\frac{V(a+h,b)+V(a-h,b)+V(a,b+h)+V(a,b-h)-4V(a,b)}{h^{2}}.
\]
This expression is equivalent to Hjorth's Laplacian estimate, which
is shown in \eqref{eq:Hjorth-approximation} with $V_{(i,j)}=V(a,b)$,
$V_{(i\pm1,j)}=V(a,b\mp h)$, and $V_{(i,j\pm1)}=V(a\pm h,b)$.

\subsection{Approximations for peripheral electrodes}

Suppose that we want to approximate $\mathrm{Lap}(V)$ at a point
$(a,b)$ which is outside the domain of $V$. Let us start with a
node in the left border of the grid, but not in a corner. Our mapping
from the continuous variables $x$ and $y$ into the discrete variables
$i$ and $j$ follows $x=(j-1)\thinspace h$ and $y=(N-i+1)\thinspace h$.
See also the scheme in Figure\,\ref{fig:xy_to_ij}. For a node on
the left border $x=0$, which corresponds to the first column of the
grid ($j=1$). The Taylor series of interest are\begin{subequations}
\begin{align*}
V(h,b) & =V(0,b)+\frac{\partial V(0,b)}{\partial x}h+\frac{1}{2}\frac{\partial^{2}V(0,b)}{\partial x^{2}}h^{2}+\frac{1}{6}\frac{\partial^{3}V(0,b)}{\partial x^{3}}h^{3}+\frac{1}{24}\frac{\partial^{4}V(0,b)}{\partial x^{4}}h^{4}+\cdots,\\
V(2h,b) & =V(0,b)+2\frac{\partial V(0,b)}{\partial x}h+2\frac{\partial^{2}V(0,b)}{\partial x^{2}}h^{2}+\frac{4}{3}\frac{\partial^{3}V(0,b)}{\partial x^{3}}h^{3}+\frac{2}{3}\frac{\partial^{4}V(0,b)}{\partial x^{4}}h^{4}+\cdots,\\
V(0,b\pm h) & =V(0,b)\pm\frac{\partial V(0,b)}{\partial y}h+\frac{1}{2}\frac{\partial^{2}V(0,b)}{\partial y^{2}}h^{2}\pm\frac{1}{6}\frac{\partial^{3}V(0,b)}{\partial y^{3}}h^{3}+\frac{1}{24}\frac{\partial^{4}V(0,b)}{\partial y^{4}}h^{4}\pm\cdots.
\end{align*}
\end{subequations}A linear combination of these expansions with weights
$-2$, 1, 1, and 1 gives 
\begin{multline*}
-2V(h,b)+V(2h,b)+V(0,b+h)+V(0,b-h)=V(0,b)+\underset{\mathrm{=Lap}(V(0,b))}{\underbrace{\left[\frac{\partial^{2}V(0,b)}{\partial x^{2}}+\frac{\partial^{2}V(0,b)}{\partial y^{2}}\right]}}h^{2}\\
+\frac{\partial^{3}V(0,b)}{\partial x^{3}}h^{3}+\cdots,
\end{multline*}
so that 
\[
\mathrm{Lap}(V(0,b))=\frac{-2V(h,b)+V(2h,b)+V(0,b+h)+V(0,b-h)-V(0,b)}{h^{2}}-\frac{\partial^{3}V(0,b)}{\partial x^{3}}h.
\]
Assuming that $h$ is sufficiently small we obtain 
\[
\mathrm{Lap}(V(0,b))\approx\frac{V(2h,b)-2V(h,b)+V(0,b+h)+V(0,b-h)-V(0,b)}{h^{2}},
\]
which is a \emph{first-order approximation}. In terms of the grid
variables,\begin{subequations}\label{eq:edges} 
\begin{equation}
\mathrm{Lap}(V)\big|_{(i,1)}\approx\frac{V_{(i,3)}-2V_{(i,2)}+V_{(i-1,1)}+V_{(i+1,1)}-V_{(i,1)}}{h^{2}},
\end{equation}
In a similar fashion, for grid points at right, bottom, and upper
edges (excluding the corners) we obtain, respectively, 
\begin{align}
\mathrm{Lap}(V)\big|_{(i,N)} & \approx\frac{V_{(i,N-2)}-2V_{(i,N-1)}+V_{(i-1,N)}+V_{(i+1,N)}-V_{(i,N)}}{h^{2}},\\
\mathrm{Lap}(V)\big|_{(N,j)} & \approx\frac{V_{(N-2,j)}-2V_{(N-1,j)}+V_{(N,j-1)}+V_{(N,j+1)}-V_{(N,,j)}}{h^{2}}\\
\mathrm{Lap}(V)\big|_{(1,j)} & \approx\frac{V_{(3,j)}-2V_{(2,j)}+V_{(1,j-1)}+V_{(1,j+1)}-V_{(1,j)}}{h^{2}}
\end{align}
\end{subequations}

Finally, assume that the node of interest is at the left upper corner
of the grid, corresponding to $i=j=1$ ($a=0$, $b=(N-1)h)$. To approximate
the Laplacian of $V$ at this node we use\begin{subequations} 
\begin{align*}
V(h,b) & =V(0,b)+\frac{\partial V(0,b)}{\partial x}h+\frac{1}{2}\frac{\partial^{2}V(0,b)}{\partial x^{2}}h^{2}+\frac{1}{6}\frac{\partial^{3}V(0,b)}{\partial x^{3}}h^{3}+\frac{1}{24}\frac{\partial^{4}V(0,b)}{\partial x^{4}}h^{4}+\cdots,\\
V(2h,b) & =V(0,b)+2\frac{\partial V(0,b)}{\partial x}h+2\frac{\partial^{2}V(0,b)}{\partial x^{2}}h^{2}+\frac{4}{3}\frac{\partial^{3}V(0,b)}{\partial x^{3}}h^{3}+\frac{2}{3}\frac{\partial^{4}V(0,b)}{\partial x^{4}}h^{4}+\cdots,\\
V(0,b-2h) & =V(0,b)-2\frac{\partial V(0,b)}{\partial y}h+2\frac{\partial^{2}V(0,b)}{\partial y^{2}}h^{2}-\frac{2}{3}\frac{\partial^{3}V(0,b)}{\partial y^{3}}h^{3}+\frac{2}{3}\frac{\partial^{4}V(0,b)}{\partial y^{4}}h^{4}+\cdots,\\
V(0,b-h) & =V(0,b)-\frac{\partial V(0,b)}{\partial y}h+\frac{1}{2}\frac{\partial^{2}V(0,b)}{\partial y^{2}}h^{2}-\frac{1}{6}\frac{\partial^{3}V(0,b)}{\partial y^{3}}h^{3}+\frac{1}{24}\frac{\partial^{4}V(0,b)}{\partial y^{4}}h^{4}+\cdots.
\end{align*}
\end{subequations}Adding these expansions with weights $-2$, $1$,
$1$, and $-2$ yields 
\begin{multline*}
-2V(h,b)+V(2h,b)+V(0,b-2h)-2V(0,b-h)=-2V(0,b)+\underset{\mathrm{=Lap}(V(a,b))}{\underbrace{\left[\frac{\partial^{2}V(0,b)}{\partial x^{2}}+\frac{\partial^{2}V(0,b)}{\partial y^{2}}\right]}}h^{2}\\
+\left[\frac{\partial^{3}V(0,b)}{\partial x^{3}}-\frac{\partial^{3}V(0,b)}{\partial y^{3}}\right]h^{3}+\cdots.
\end{multline*}
Solving for $\mathrm{Lap}(V)$ we obtain the first-order approximation
\[
\mathrm{Lap}(V(0,b))\approx\frac{V(2h,b)-2V(h,b)+V(0,b-2h)-2V(0,b-h)+2V(0,b)}{h^{2}}.
\]
Equivalently, 
\begin{equation}
\mathrm{Lap}(V)\big|_{(1,1)}\approx\frac{V_{(1,3)}-2V_{(1,2)}+V_{(3,1)}-2V_{(2,1)}+2V_{(1,1)}}{h^{2}}.
\end{equation}
Using symmetry we can obtain similar approximations for the other
corners of the grid illustrated in Figure\,\ref{fig:Five-point-stencils}.

\section{A Matlab Code to Construct $\mathbf{S_{\lambda}}$ and $\mathbf{L}_{\lambda}$\label{sec:Matlab-Code-For-Spherical-Splines}}

Here we present a Matlab code to build the smoother matrix $\mathbf{S_{\lambda}}$
and the surface Laplacian matrix $\mathbf{L}_{\lambda}$ using spherical
splines. The first function, \texttt{sphlap0.m,} is an auxiliary function
to generate the matrices $\mathbf{K}$, $\tilde{\mathbf{K}}$, $\mathbf{T}$,
$\mathbf{Q}_{1}$, $\mathbf{Q}_{2}$, and $\mathbf{R}$ presented
above. Since none of these matrices depend on the regularization parameter
$\lambda$, they need to be computed only once for each value of $m$
and electrode configuration. The computation of these matrices is
not fast because it involves the evaluation of Legendre polynomials.
Therefore, for efficiency, this function should never be called inside
a regularization loop. The second function, \texttt{sphlap.m,} receives
the output \texttt{of sphlap0.m} as argument plus the value of $\lambda$
and return the matrices $\mathbf{S_{\lambda}}$ and $\mathbf{L}_{\lambda}$.
This is the function that should be used inside a regularization loop.
\medskip{}

\noindent 
\begin{lstlisting}[basicstyle={\small\ttfamily},basicstyle={\ttfamily\linespread{0.85}\footnotesize},breaklines=true,caption={Auxiliar function to compute spherical splines },columns=fullflexible,language=Matlab,showstringspaces=false,tabsize=4]
% sphlap0.m - Auxiliar function to compute spherical splines
%
% Usage: [K, LapK, T, Q1, Q2, R] = sphlap0(x, y, z, m, tol);
%
% Required Inputs
%   x, y, z: electrode coordinates (must be on a sphere)
%   m: interpolation order (2<=m<=6)
%   lambda: smoothing parameter
% Optional Input
%   tol: error tolerance in the estimate of the Legendre polynomials. 
%        Default is 1e-10.
%
% Output
%   K, LapK, T, Q1, Q2, R: matrices required to implement spherical splines
%

function [K, LapK, T, Q1, Q2, R, tol] = sphlap0(x, y, z, m, tol)

% Handle arguments
if nargin < 4
	help sphlap0.m;
	return;
end

if nargin == 4
    tol = 1e-10;
end

if m < 2 || m > 6
	error('sphlap0:intorder','The parameter "m" should be in the range of %s',...
        '2 and 6');
end

r = hypot(z, hypot(x, y)); % head radius
N = length(x); % number of electrodes

sqdist = pdist([x(:) y(:) z(:)], 'euclidean').^2; % square distances between electrodes
dr2 = squareform(sqdist); % convert distances to matrix form
r2 = r(1)^2; % squared head radius
cos_gamma = 1 - dr2/(2*r2);	% angle between electrodes

if any((cos_gamma(:) > 1) | (cos_gamma(:) < -1))
    error('sphlap0:locs', ...
        'Something is wrong with the electrode coordinates. %s', ...
        'Are they on located on a sphere');
end

G = [];
LapG = [];
G0 = 0;
epsilon = tol + 1;
n = 1;
while (tol < epsilon)
	Pn = legendre(n, cos_gamma(:));
    a = (2*n+1) / (n*(n+1))^m;
    gm = a * Pn(1,:)';
    G = horzcat(G, gm);
    LapG = horzcat(LapG, -n * (n+1) * gm);
    epsilon = max(abs(G(:,end)-G0));
    G0 = G(:,end);
    n = n + 1;
end

tol = epsilon; % final error tolerance

K = reshape(sum(G,2), N, [])/(4*pi);
LapK = reshape(sum(LapG,2), N, [])/(4*pi*r2);
T = ones(N,1);

% QR decomposition of T
[Q, R] = qr(T);
R = R(1);
Q1 = Q(:,1);
Q2 = Q(:,2:N);

% Alternative
% R = -sqrt(N);
% Q1 = T / R;
% [U,~,~] = svd(T);
% Q2 = U(:,2:end);

end
\end{lstlisting}

\noindent 
\begin{lstlisting}[basicstyle={\ttfamily\linespread{0.85}\footnotesize},breaklines=true,caption={Function to compute the smoother and the Laplacian matrices },columns=fullflexible,language=Matlab,showstringspaces=false,tabsize=4]
% sphlap.m - Compute the smoother and Laplacian matrices using spherical splines
%
% Usage: [S, L] = sphlap(K, LapK, T, Q1, Q2, R, lambda);
%
% Required Inputs
%   K, LapK, T, Q1, Q2, and R: matrices generated by sphlap0
%	lambda: smoothing parameter
%
% Output:
%   S = the smoother matrix
%   L = the Laplacian matrix
%

function [S, L] = sphlap(K, LapK, T, Q1, Q2, R, lambda)

% Handle arguments
if nargin ~= 7
	help sphlap.m;
	return;
end

I = eye(size(K,1));
Klamb = K + lambda*I;

C = Q2 / (Q2' * Klamb * Q2) * Q2';
D = R \ Q1' * (I - Klamb*C);

S = K * C + T * D; % The smoother matrix
L = LapK * C; % The Laplacian matrix

end
\end{lstlisting}
\bibliographystyle{plainnat}
\bibliography{IJP}

\begin{thebibliography}{80}
\providecommand{\natexlab}[1]{#1}
\providecommand{\url}[1]{\texttt{#1}}
\expandafter\ifx\csname urlstyle\endcsname\relax
  \providecommand{\doi}[1]{doi: #1}\else
  \providecommand{\doi}{doi: \begingroup \urlstyle{rm}\Url}\fi

\bibitem[Abramowitz and Stegun(1964)]{Abramowitz1964}
Milton Abramowitz and Irene~A. Stegun.
\newblock \emph{Handbook of mathematical functions: with formulas, graphs, and
  mathematical tables}.
\newblock Dover Publications, Inc., New York, 1964.

\bibitem[Alschuler et~al.(2014)Alschuler, Tenke, Bruder, and
  Kayser]{Alschuler2014}
Daniel~M. Alschuler, Craig~E. Tenke, Gerard~E. Bruder, and J{\"u}rgen Kayser.
\newblock Identifying electrode bridging from electrical distance
  distributions: a survey of publicly-available {EEG} data using a new method.
\newblock \emph{Clin. Neurophysiol.}, 125:\penalty0 484--490, 2014.
\newblock \doi{10.1016/j.clinph.2013.08.024}.

\bibitem[Astolfi et~al.(2007)Astolfi, Cincotti, Mattia, Marciani, Baccala, {de
  Vico Fallani}, Salinari, Ursino, Zavaglia, Ding, Edgar, Miller, He, and
  Babiloni]{Astolfi2007}
Laura Astolfi, Febo Cincotti, Donatella Mattia, M~Grazia Marciani, Luiz~A.
  Baccala, Fabrizio {de Vico Fallani}, Serenella Salinari, Mauro Ursino,
  Melissa Zavaglia, Lei Ding, J~Christopher Edgar, Gregory~A. Miller, Bin He,
  and Fabio Babiloni.
\newblock Comparison of different cortical connectivity estimators for
  high-resolution {EEG} recordings.
\newblock \emph{Hum. Brain Mapp.}, 28:\penalty0 143--157, 2007.
\newblock \doi{10.1002/hbm.20263}.

\bibitem[Babiloni et~al.(1995)Babiloni, Babiloni, Fattorini, Carducci, Onorati,
  and Urbano]{babiloni_performances_1995}
F.~Babiloni, C.~Babiloni, L.~Fattorini, F.~Carducci, P.~Onorati, and A.~Urbano.
\newblock Performances of surface {L}aplacian estimators: A study of simulated
  and real scalp potential distributions.
\newblock \emph{Brain Topogr.}, 8:\penalty0 35--45, 1995.

\bibitem[Babiloni et~al.(1996)Babiloni, Babiloni, Carducci, Fattorini, Onorati,
  and Urbano]{babilon_spline_Laplacian_1996}
F.~Babiloni, C.~Babiloni, F.~Carducci, L.~Fattorini, P.~Onorati, and A.~Urbano.
\newblock Spline {L}aplacian estimate of {EEG} potentials over a realistic
  magnetic resonance-contructed scalp surface model.
\newblock \emph{Electroencephalogr. Clin. Neurophysiol.}, 98:\penalty0
  363--373, 1996.

\bibitem[Boneva et~al.(1971)Boneva, Kendall, and Stefanov]{Boneva1971}
L.~I. Boneva, D.~G. Kendall, and I.~Stefanov.
\newblock Spline transformations: {Three} new diagnostic aids for statistical
  data-analyst.
\newblock \emph{J. Royal Statist. Soc. Ser. B}, 33:\penalty0 1--70, 1971.

\bibitem[Bortel and Sovka(2007)]{Bortel2007}
Radoslav Bortel and Pavel Sovka.
\newblock Regularization techniques in realistic laplacian computation.
\newblock \emph{IEEE Trans Biomed Eng}, 54\penalty0 (11):\penalty0 1993--1999,
  Nov 2007.
\newblock \doi{10.1109/TBME.2007.893496}.
\newblock URL \url{http://dx.doi.org/10.1109/TBME.2007.893496}.

\bibitem[Bortel and Sovka(2008)]{Bortel2008}
Radoslav Bortel and Pavel Sovka.
\newblock Electrode position scaling in realistic laplacian computation.
\newblock \emph{IEEE Trans Biomed Eng}, 55\penalty0 (9):\penalty0 2314--2316,
  Sep 2008.
\newblock \doi{10.1109/TBME.2008.921168}.
\newblock URL \url{http://dx.doi.org/10.1109/TBME.2008.921168}.

\bibitem[Bortel and Sovka(2013)]{Bortel2013}
Radoslav Bortel and Pavel Sovka.
\newblock Potential approximation in realistic laplacian computation.
\newblock \emph{Clin Neurophysiol}, 124\penalty0 (3):\penalty0 462--473, Mar
  2013.
\newblock \doi{10.1016/j.clinph.2012.08.020}.
\newblock URL \url{http://dx.doi.org/10.1016/j.clinph.2012.08.020}.

\bibitem[Carvalhaes et~al.(2014)Carvalhaes, {de Barros}, Perreau-Guimaraes, and
  Suppes]{Carvalhaes2014}
C.~G. Carvalhaes, J~Acacio {de Barros}, M.~Perreau-Guimaraes, and P.~Suppes.
\newblock The joint use of the tangential electric field and surface
  {L}aplacian in {EEG} classification.
\newblock \emph{Brain Topogr.}, 27:\penalty0 84--94, 2014.
\newblock \doi{10.1007/s10548-013-0305-y}.

\bibitem[Carvalhaes et~al.(2009)Carvalhaes, Perreau-Guimaraes, Grosenick, and
  Suppes]{Carvalhaes2009}
C.G. Carvalhaes, M.~Perreau-Guimaraes, L.~Grosenick, and P.~Suppes.
\newblock {EEG} classification by {ICA} source selection of
  {L}aplacian-filtered data.
\newblock In \emph{Proc. IEEE ISBI 09}, pages 1003--1006, 2009.

\bibitem[Carvalhaes(2013)]{Carvalhaes2013}
Claudio Carvalhaes.
\newblock Spline interpolation on nonunisolvent sets.
\newblock \emph{IMA J. Num. Anal.}, 33:\penalty0 370--375, 2013.

\bibitem[Carvalhaes and Suppes(2011)]{Carvalhaes2011}
Claudio Carvalhaes and Patrick Suppes.
\newblock A spline framework for estimating the {EEG} surface laplacian using
  the {E}uclidean metric.
\newblock \emph{Neural Comput.}, 23\penalty0 (11):\penalty0 2974--3000, Nov
  2011.

\bibitem[Chi et~al.(2010)Chi, Jung, and Cauwenberghs]{chi2010dry}
Yu~Mike Chi, Tzyy-Ping Jung, and Gert Cauwenberghs.
\newblock Dry-contact and noncontact biopotential electrodes: methodological
  review.
\newblock \emph{Biomedical Engineering, IEEE Reviews in}, 3:\penalty0 106--119,
  2010.

\bibitem[Craven and Wahba(1979)]{Craven1979}
P~Craven and G~Wahba.
\newblock Smoothing noisy data with spline functions: {E}stimating the correct
  degree of smoothing by the method of generalized cross-validation.
\newblock \emph{Numer Math}, pages 377--403, 1979.

\bibitem[de~Barros et~al.(2006)de~Barros, Carvalhaes, de~Mendon\c{c}a, and
  Suppes]{Barros2006}
J~Acacio de~Barros, ClaudiClaudio Carvalhaes, J.~P. R.~F. de~Mendon\c{c}a, and
  P.~Suppes.
\newblock Recognition of words from the {EEG} {L}aplacian.
\newblock \emph{Braz. J. Biom. Eng.}, 21:\penalty0 45--59, 2006.

\bibitem[{Del Percio} et~al.(2007){Del Percio}, Brancucci, Bergami, Marzano,
  Fiore, {Di Ciolo}, Aschieri, Lino, Vecchio, Iacoboni, Gallamini, Babiloni,
  and Eusebi]{DelPercio2007}
Claudio {Del Percio}, Alfredo Brancucci, Francesca Bergami, Nicola Marzano,
  Antonio Fiore, Enrico {Di Ciolo}, Pierluigi Aschieri, Andrea Lino, Fabrizio
  Vecchio, Marco Iacoboni, Michele Gallamini, Claudio Babiloni, and Fabrizio
  Eusebi.
\newblock Cortical alpha rhythms are correlated with body sway during quiet
  open-eyes standing in athletes: a high-resolution {EEG} study.
\newblock \emph{Neuroimage}, 36:\penalty0 822--829, 2007.
\newblock \doi{10.1016/j.neuroimage.2007.02.054}.

\bibitem[Delorme and Makeig(2004)]{Delorme2004}
Arnaud Delorme and Scott Makeig.
\newblock {EEGLAB}: an open source toolbox for analysis of single-trial {EEG}
  dynamics including independent component analysis.
\newblock \emph{J. Neurosci. Meth.}, 134:\penalty0 9--21, 2004.
\newblock \doi{10.1016/j.jneumeth.2003.10.009}.

\bibitem[Doesburg et~al.(2008)Doesburg, Roggeveen, Kitajo, and
  Ward]{Doesburg2008}
Sam~M. Doesburg, Alexa~B. Roggeveen, Keiichi Kitajo, and Lawrence~M. Ward.
\newblock Large-scale gamma-band phase synchronization and selective attention.
\newblock \emph{Cereb. Cortex}, 18:\penalty0 386--396, 2008.
\newblock \doi{10.1093/cercor/bhm073}.

\bibitem[Duchon(1977)]{duchon_splines_1977}
J.~Duchon.
\newblock Splines minimizing rotation-invariant semi-norms in {S}obolev spaces.
\newblock In \emph{Constructive theory of functions of several variables},
  volume 571, pages 85--100, 1977.

\bibitem[Eubank(1999)]{Eubank:1999jk}
Randall~L Eubank.
\newblock \emph{Nonparametric regression and spline smoothing}, volume 157 of
  \emph{Statistics: textbooks and monographs}.
\newblock Marcel Dekker, New York, 2nd edition, 1999.

\bibitem[Fitzgibbon et~al.(2013)Fitzgibbon, Lewis, Powers, Whitham, Willoughby,
  and Pope]{Fitzgibbon2013}
S~P Fitzgibbon, T~W Lewis, D~M Powers, E~W Whitham, J~O Willoughby, and K~J
  Pope.
\newblock Surface {L}aplacian of central scalp electrical signals is
  insensitive to muscle contamination.
\newblock \emph{IEEE Trans. Biom. Eng.}, 60:\penalty0 4--9, 2013.

\bibitem[Freeman and Nicholson(1975)]{Freeman1975}
J.~A. Freeman and C.~Nicholson.
\newblock Experimental optimization of current source-density technique for
  anuran cerebellum.
\newblock \emph{J. Neurophysiol.}, 38:\penalty0 369--382, 1975.

\bibitem[Gevins(1988)]{Gevins1988}
A.~Gevins.
\newblock \emph{Dynamics of Sensory and Cognitive Processing of the Brain},
  chapter Recent advances in neurocognitive pattern analysis, pages 88--102.
\newblock Springer-Verlag, 1988.

\bibitem[Gevins(1989)]{gevins_dynamic_1989}
A.~Gevins.
\newblock Dynamic functional topography of cognitive tasks.
\newblock \emph{Brain Topogr.}, 2:\penalty0 37--56, 1989.

\bibitem[Gevins et~al.(1990)Gevins, Brickett, Costales, Le, and
  Reutter]{Gevins1990}
A.~Gevins, P.~Brickett, B.~Costales, J.~Le, and B.~Reutter.
\newblock Beyond topographic mapping: towards functional-anatomical imaging
  with 124-channel {EEG}s and 3-{D} {MRI}s.
\newblock \emph{Brain Topogr.}, 3:\penalty0 53--64, 1990.

\bibitem[Gevins et~al.(1994)Gevins, Le, Martin, Brickett, Desmond, and
  Reutter]{gevins_high_1994}
A.~Gevins, J.~Le, N.~K. Martin, P.~Brickett, J.~Desmond, and B.~Reutter.
\newblock High resolution {EEG:} 124-channel recording, spatial deblurring and
  {MRI} integration methods.
\newblock \emph{Electroencephalogr. Clin. Neurophysiol.}, 90:\penalty0 337--58,
  1994.

\bibitem[Green and Silverman(1994)]{Green1994}
P.~J. Green and B.~W. Silverman.
\newblock \emph{Nonparametric regression and generalized linear models: a
  roughness penalty approach}.
\newblock Monographs on Statistics \& Applied Probability. Chapman \& Hall,
  {L}ondon, 1994.

\bibitem[Greischar et~al.(2004)Greischar, Burghy, {van Reekum}, Jackson,
  Pizzagalli, Mueller, and Davidson]{Greischar2004}
Lawrence~L. Greischar, Cory~A. Burghy, Carien~M. {van Reekum}, Daren~C.
  Jackson, Diego~A. Pizzagalli, Corrina Mueller, and Richard~J. Davidson.
\newblock Effects of electrode density and electrolyte spreading in dense array
  electroencephalographic recording.
\newblock \emph{Clin. Neurophysiol.}, 115:\penalty0 710--720, 2004.
\newblock \doi{10.1016/j.clinph.2003.10.028}.

\bibitem[H\"am\"al\"ainen et~al.(1993)H\"am\"al\"ainen, Hari, Ilmoniemi,
  Knuutila, and Lounasmaa]{Haemaelaeinen1993}
Matti H\"am\"al\"ainen, Riitta Hari, Risto~J. Ilmoniemi, Jukka Knuutila, and
  Olli~V. Lounasmaa.
\newblock Magnetoencephalography\char22{}theory, instrumentation, and
  applications to noninvasive studies of the working human brain.
\newblock \emph{Rev. Mod. Phys.}, 65:\penalty0 413--497, 1993.

\bibitem[Hastie et~al.(2009)Hastie, Tibshirani, Friedman, Hastie, Friedman, and
  Tibshirani]{Hastie2009}
Trevor Hastie, Robert Tibshirani, Jerome Friedman, T~Hastie, J~Friedman, and
  R~Tibshirani.
\newblock \emph{The Elements of Statistical Learning: Data Mining, Inference,
  and Prediction}.
\newblock Springer Series in Statistics. Springer, New York, 2009.

\bibitem[Hjorth(1975)]{hjorth_-line_1975}
B.~Hjorth.
\newblock An on-line transformation of {EEG} scalp potentials into orthogonal
  source derivations.
\newblock \emph{Electroencephalogr. Clin. Neurophysiol.}, 39:\penalty0
  526--530, 1975.

\bibitem[Huigen et~al.(2002)Huigen, Peper, and
  Grimbergen]{huigen2002investigation}
E~Huigen, A~Peper, and CA~Grimbergen.
\newblock Investigation into the origin of the noise of surface electrodes.
\newblock \emph{Medical and biological engineering and computing}, 40\penalty0
  (3):\penalty0 332--338, 2002.

\bibitem[Jackson(1999)]{jackson_classical_1999}
John~David Jackson.
\newblock \emph{Classical Electrodynamics, 3rd ed.}
\newblock Wiley, New York, 1999.

\bibitem[James et~al.(2013)James, Witten, Hastie, and Tibshirani]{James2013}
Gareth James, Daniela Witten, Trevor Hastie, and Robert Tibshirani.
\newblock \emph{An Introduction to Statistical Learning with Applications in
  {R}}.
\newblock Springer Texts in Statistics. Springer, New York, 2013.

\bibitem[Kayser and Tenke(2006{\natexlab{a}})]{Kayser2006}
J{\"u}rgen Kayser and Craig~E Tenke.
\newblock Principal components analysis of {L}aplacian waveforms as a generic
  method for identifying {ERP} generator patterns: {II}. adequacy of
  low-density estimates.
\newblock \emph{Clin. Neurophysiol.}, 117:\penalty0 369--380,
  2006{\natexlab{a}}.

\bibitem[Kayser and Tenke(2006{\natexlab{b}})]{Kayser2006a}
J{\"u}rgen Kayser and Craig~E Tenke.
\newblock Principal components analysis of {L}aplacian waveforms as a generic
  method for identifying {ERP} generator patterns: {I}. evaluation with
  auditory oddball tasks.
\newblock \emph{Clin. Neurophysiol.}, 117:\penalty0 348--368,
  2006{\natexlab{b}}.

\bibitem[Law et~al.(1993{\natexlab{a}})Law, Nunez, and
  Wijesinghe]{law_high-resolution_1993}
S.~K. Law, P.~L. Nunez, and R.~S. Wijesinghe.
\newblock High-resolution {EEG} using spline generated surface {L}aplacians on
  spherical and ellipsoidal surfaces.
\newblock \emph{{IEEE} Trans. Biom. Eng.}, 40:\penalty0 145--153,
  1993{\natexlab{a}}.

\bibitem[Law et~al.(1993{\natexlab{b}})Law, Rohrbaugh, Adams, and
  Eckardt]{law_improving_1993}
S.~K. Law, J.~W. Rohrbaugh, C.~M. Adams, and M.~J. Eckardt.
\newblock Improving spatial and temporal resolution in evoked {EEG} responses
  using surface {L}aplacians.
\newblock \emph{Electroencephalogr. Clin. Neurophysiol.}, 88:\penalty0
  309--322, 1993{\natexlab{b}}.

\bibitem[Le and Gevins(1993)]{Le1993}
J.~Le and A.~Gevins.
\newblock Method to reduce blur distortion from {EEG}'s using a realistic head
  model.
\newblock \emph{{IEEE} Trans. Biom. Eng.}, 40:\penalty0 517--528, 1993.

\bibitem[Le et~al.(1994)Le, Menon, and Gevins]{le_local_1994}
J.~Le, V.~Menon, and A.~Gevins.
\newblock Local estimate of surface {L}aplacian derivation on a realistically
  shaped scalp surface and its performance on noisy data.
\newblock \emph{Electroencephalogr. Clin. Neurophysiol.}, 92:\penalty0
  433--441, 1994.

\bibitem[Lu et~al.(2013)Lu, McFarland, and Wolpaw]{Lu2013}
Jun Lu, Dennis~J. McFarland, and Jonathan~R. Wolpaw.
\newblock Adaptive laplacian filtering for sensorimotor rhythm-based
  brain-computer interfaces.
\newblock \emph{J Neural Eng}, 10\penalty0 (1):\penalty0 016002, Feb 2013.
\newblock \doi{10.1088/1741-2560/10/1/016002}.
\newblock URL \url{http://dx.doi.org/10.1088/1741-2560/10/1/016002}.

\bibitem[McFarland et~al.(1997)McFarland, McCane, David, and
  Wolpaw]{McFarland1997}
D.~J. McFarland, L.~M. McCane, S.~V. David, and J.~R. Wolpaw.
\newblock Spatial filter selection for {EEG}-based communication.
\newblock \emph{Electroencephalogr. Clin. Neurophysiol.}, 103:\penalty0
  386--394, 1997.

\bibitem[Meinguet(1979)]{meinguet_1979}
Jean Meinguet.
\newblock Multivariate interpolation at arbitrary points made simple.
\newblock \emph{J. App. Math. Phys.}, 30:\penalty0 292--304, 1979.

\bibitem[{Metting van Rijn} et~al.(1990){Metting van Rijn}, Peper, and
  Grimbergen]{MettingvanRijn1990}
A.~C. {Metting van Rijn}, A.~Peper, and C.~A. Grimbergen.
\newblock High-quality recording of bioelectric events. part 1. interference
  reduction, theory and practice.
\newblock \emph{Med. Biol. Eng. Comput.}, 28:\penalty0 389--397, 1990.

\bibitem[{Metting van Rijn} et~al.(1991){Metting van Rijn}, Peper, and
  Grimbergen]{MettingvanRijn1991}
A.~C. {Metting van Rijn}, A.~Peper, and C.~A. Grimbergen.
\newblock High-quality recording of bioelectric events. part 2. low-noise,
  low-power multichannel amplifier design.
\newblock \emph{Med. Biol. Eng. Comput.}, 29:\penalty0 433--440, 1991.

\bibitem[M\"{u}ller-Gerking et~al.(1999)M\"{u}ller-Gerking, Pfurtscheller, and
  Flyvbjerg]{Mueller-Gerking1999}
J.~M\"{u}ller-Gerking, G.~Pfurtscheller, and H.~Flyvbjerg.
\newblock Designing optimal spatial filters for single-trial {EEG}
  classification in a movement task.
\newblock \emph{Clin. Neurophysiol.}, 110:\penalty0 787--798, 1999.

\bibitem[Nicholson(1973)]{nicholson_theoretical_1973}
C.~Nicholson.
\newblock Theoretical analysis of field potentials in anisotropic ensembles of
  neuronal elements.
\newblock \emph{IEEE Trans. Biomed. Eng.}, 20:\penalty0 278--288, 1973.
\newblock \doi{10.1109/TBME.1973.324192}.

\bibitem[Nicholson and Freeman(1975)]{nicholson_theory_1975}
C.~Nicholson and J.~A. Freeman.
\newblock Theory of current source-density analysis and determination of
  conductivity tensor for anuran cerebellum.
\newblock \emph{J. Neurophysiol.}, 38:\penalty0 356--368, 1975.

\bibitem[Nunez and Srinivasan(2006)]{nunez_electric_2006}
P.~L. Nunez and R.~Srinivasan.
\newblock \emph{Electric Fields of the Brain: The Neurophysics of {EEG}}.
\newblock Oxford University Press, New York, 2nd edition, 2006.

\bibitem[Nunez et~al.(1993)Nunez, Silberstein, Cadusch, and
  Wijesinghe]{nunez_comparison_1993}
P.~L. Nunez, R.~B. Silberstein, P.~J. Cadusch, and R.~Wijesinghe.
\newblock Comparison of high resolution {EEG} methods having different
  theoretical bases.
\newblock \emph{Brain Topogr.}, 5:\penalty0 361--364, 1993.

\bibitem[Nunez et~al.(1994)Nunez, Silberstein, Cadusch, Wijesinghe, Westdorp,
  and Srinivasan]{nunez_theoretical_1994}
P.~L. Nunez, R.~B. Silberstein, P.~J. Cadusch, R.~S. Wijesinghe, A.~F.
  Westdorp, and R.~Srinivasan.
\newblock A theoretical and experimental study of high resolution {EEG} based
  on surface laplacians and cortical imaging.
\newblock \emph{Electroencephalogr. Clin. Neurophysiol.}, 90:\penalty0 40--57,
  1994.

\bibitem[Nunez and Westdorp(1994)]{nunez1994}
Paul~L. Nunez and Andrew~F. Westdorp.
\newblock The surface {L}aplacian, high-resolution {EEG} and controversies.
\newblock \emph{Brain Topogr.}, 6:\penalty0 221--226, 1994.

\bibitem[Nunez et~al.(2001)Nunez, Wingeier, and Silberstein]{nunez_2001}
Paul~L. Nunez, Brett~M. Wingeier, and Richard~B. Silberstein.
\newblock Spatial-temporal structures of human alpha rhythms: Theory,
  microcurrent sources, multiscale measurements, and global binding of local
  networks.
\newblock \emph{Hum. Brain Mapp.}, 13:\penalty0 125--164, 2001.

\bibitem[Perrin et~al.(1987{\natexlab{a}})Perrin, Bertrand, and
  Pernier]{perrin_scalp_1987}
F.~Perrin, O.~Bertrand, and J.~Pernier.
\newblock Scalp current density mapping: value and estimation from potential
  data.
\newblock \emph{{IEEE} Trans Biomed Eng}, 34:\penalty0 283--288,
  1987{\natexlab{a}}.

\bibitem[Perrin et~al.(1987{\natexlab{b}})Perrin, Pernier, Bertrand, Giard, and
  Echallier]{perrin_mapping_1987}
F.~Perrin, J.~Pernier, O.~Bertrand, M.~H. Giard, and J.~F. Echallier.
\newblock Mapping of scalp potentials by surface spline interpolation.
\newblock \emph{Electroencephalogr Clin Neurophysiol}, 66:\penalty0 75--81,
  1987{\natexlab{b}}.

\bibitem[Perrin et~al.(1989)Perrin, Pernier, Bertrand, and
  Echallier]{perrin_spherical_1989}
F.~Perrin, J.~Pernier, O.~Bertrand, and J.~F. Echallier.
\newblock Spherical splines for scalp potential and current density mapping.
\newblock \emph{Electroencephalogr. Clin. Neurophysiol.}, 72:\penalty0
  184--187, 1989.

\bibitem[Petrov(2012)]{petrov2012anisotropic}
Yury Petrov.
\newblock Anisotropic spherical head model and its application to imaging
  electric activity of the brain.
\newblock \emph{Physical Review E}, 86\penalty0 (1):\penalty0 011917, 2012.

\bibitem[Press(1992)]{press_numerical_1992}
W.~H. Press.
\newblock \emph{Numerical Recipes in C: the art of scientific computing}.
\newblock Cambridge University Press, 1992.

\bibitem[Rush and Driscoll(1968)]{rush1968current}
Stanley Rush and Daniel~A Driscoll.
\newblock Current distribution in the brain from surface electrodes.
\newblock \emph{Anesthesia \& Analgesia}, 47\penalty0 (6):\penalty0 717--723,
  1968.

\bibitem[Schey(2004)]{schey_div_2004}
H.~M. Schey.
\newblock \emph{Div, Grad, Curl, And All That: An Informal Text On Vector
  Calculus}.
\newblock W. W. Norton \& Company, 2004.

\bibitem[Sibson and Stone(1991)]{Sibson1991}
Robin Sibson and G.~Stone.
\newblock Computation of thin-plate splines.
\newblock \emph{SIAM J. Sci. Stat. Comput.}, 12:\penalty0 1304--1313, 1991.

\bibitem[Spinelli and Haberman(2010)]{spinelli2010insulating}
Enrique Spinelli and Marcelo Haberman.
\newblock Insulating electrodes: a review on biopotential front ends for
  dielectric skin--electrode interfaces.
\newblock \emph{Physiological measurement}, 31\penalty0 (10):\penalty0 S183,
  2010.

\bibitem[Srinivasan et~al.(2007)Srinivasan, Winter, Ding, and
  Nunez]{Srinivasan2007}
Ramesh Srinivasan, William~R. Winter, Jian Ding, and Paul~L. Nunez.
\newblock {EEG} and {MEG} coherence: measures of functional connectivity at
  distinct spatial scales of neocortical dynamics.
\newblock \emph{J. Neurosci. Meth.}, 166:\penalty0 41--52, 2007.
\newblock \doi{10.1016/j.jneumeth.2007.06.026}.

\bibitem[Tenke and Kayser(2001)]{Tenke2001}
C.~E. Tenke and J.~Kayser.
\newblock A convenient method for detecting electrolyte bridges in multichannel
  electroencephalogram and event-related potential recordings.
\newblock \emph{Clin. Neurophysiol.}, 112:\penalty0 545--550, 2001.

\bibitem[Tenke et~al.(1993)Tenke, Schroeder, Arezzo, and Vaughan]{Tenke1993}
C.~E. Tenke, C.~E. Schroeder, J.~C. Arezzo, and HG~Vaughan, Jr.
\newblock Interpretation of high-resolution current source density profiles: a
  simulation of sublaminar contributions to the visual evoked potential.
\newblock \emph{Exp. Brain Res.}, 94:\penalty0 183--192, 1993.

\bibitem[Tenke et~al.(1998)Tenke, Kayser, Fong, Leite, Towey, and
  Bruder]{Tenke1998}
C.~E. Tenke, J.~Kayser, R.~Fong, P.~Leite, J.~P. Towey, and G.~E. Bruder.
\newblock Response- and stimulus-related {ERP} asymmetries in a tonal oddball
  task: a {L}aplacian analysis.
\newblock \emph{Brain Topogr.}, 10:\penalty0 201--210, 1998.

\bibitem[Tenke and Kayser(2012)]{Tenke2012}
Craig~E. Tenke and J{\"u}rgen Kayser.
\newblock Generator localization by current source density ({CSD}):
  implications of volume conduction and field closure at intracranial and scalp
  resolutions.
\newblock \emph{Clin. Neurophysiol.}, 123:\penalty0 2328--2345, 2012.
\newblock \doi{10.1016/j.clinph.2012.06.005}.

\bibitem[Tenke et~al.(2011)Tenke, Kayser, Manna, Fekri, Kroppmann, Schaller,
  Alschuler, Stewart, McGrath, and Bruder]{Tenke2011}
Craig~E. Tenke, Jürgen Kayser, Carlye~G. Manna, Shiva Fekri, Christopher~J.
  Kroppmann, Jennifer~D. Schaller, Daniel~M. Alschuler, Jonathan~W. Stewart,
  Patrick~J. McGrath, and Gerard~E. Bruder.
\newblock Current source density measures of electroencephalographic alpha
  predict antidepressant treatment response.
\newblock \emph{Biol Psychiatry}, 70\penalty0 (4):\penalty0 388--394, Aug 2011.
\newblock \doi{10.1016/j.biopsych.2011.02.016}.
\newblock URL \url{http://dx.doi.org/10.1016/j.biopsych.2011.02.016}.

\bibitem[Wahba(1976)]{Wahba1976}
G~Wahba.
\newblock Histosplines with knots which are order statistics.
\newblock \emph{J. R. Stat. Soc. B}, 38:\penalty0 140--151, 1976.

\bibitem[Wahba(1981)]{Wahba1981}
G.~Wahba.
\newblock Numerical experiments with the thin plate histospline.
\newblock \emph{Comm. Statist. - Theor. Meth.}, 10:\penalty0 2475--2514, 1981.

\bibitem[Wahba(1990)]{wahba_spline_1990}
G.~Wahba.
\newblock \emph{Spline models for observational data}, volume~59 of
  \emph{{CBMS-NSF} Regional Conference Series in Applied Mathematics}.
\newblock {SIAM}, Philadelphia, 1990.

\bibitem[Wang et~al.(2012)Wang, Perreau-Guimaraes, Carvalhaes, and
  Suppes]{Wang2012}
Rui Wang, Marcos Perreau-Guimaraes, Claudio Carvalhaes, and Patrick Suppes.
\newblock Using phase to recognize english phonemes and their distinctive
  features in the brain.
\newblock \emph{Proc. Natl. Acad. Sci.}, 109:\penalty0 20685--20690, 2012.
\newblock \doi{10.1073/pnas.1217500109}.

\bibitem[Winter et~al.(2007)Winter, Nunez, Ding, and Srinivasan]{Winter2007}
William~R. Winter, Paul~L. Nunez, Jian Ding, and Ramesh Srinivasan.
\newblock Comparison of the effect of volume conduction on {EEG} coherence with
  the effect of field spread on {MEG} coherence.
\newblock \emph{Stat. Med.}, 26:\penalty0 3946--3957, 2007.
\newblock \doi{10.1002/sim.2978}.

\bibitem[Wolters et~al.(2006)Wolters, Anwander, Tricoche, Weinstein, Koch, and
  MacLeod]{wolters2006influence}
CH~Wolters, A~Anwander, X~Tricoche, D~Weinstein, MA~Koch, and RS~MacLeod.
\newblock Influence of tissue conductivity anisotropy on eeg/meg field and
  return current computation in a realistic head model: a simulation and
  visualization study using high-resolution finite element modeling.
\newblock \emph{NeuroImage}, 30\penalty0 (3):\penalty0 813--826, 2006.

\bibitem[Wood(2003)]{Wood2003}
S.~N. Wood.
\newblock Thin plate regression splines.
\newblock \emph{J. R. Statist. Soc. B}, pages 95--114, 2003.

\bibitem[Yao(1996)]{Yao1996}
Yao.
\newblock The equivalent source technique and cortical imaging.
\newblock \emph{Electroencephalogr. Clin. Neurophysiol.}, 98:\penalty0
  478--483, 1996.

\bibitem[Yao(2002)]{yao_high-resolution_2002}
D.~Yao.
\newblock High-resolution {EEG} mapping: a radial-basis function based approach
  to the scalp laplacian estimate.
\newblock \emph{Clin. Neurophysiol.}, 113:\penalty0 956--967, 2002.

\bibitem[Yao et~al.(2001)Yao, Zhou, Zeng, Fan, Lian, Wu, Ao, Chen, and
  He]{Yao2001}
D.~Yao, Y.~Zhou, M.~Zeng, S.~Fan, J.~Lian, D.~Wu, X.~Ao, L.~Chen, and B.~He.
\newblock A study of equivalent source techniques for high-resolution {EEG}
  imaging.
\newblock \emph{Phys. Med. Biol.}, 46:\penalty0 2255--2266, 2001.

\bibitem[Zhai and Yao(2004)]{Zhai2004}
Yiran Zhai and Dezhong Yao.
\newblock A radial-basis function based surface {L}aplacian estimate for a
  realistic head model.
\newblock \emph{Brain Topogr.}, 17:\penalty0 55--62, 2004.

\end{thebibliography}

\end{document}